\documentclass[a4paper,11pt]{article}
\pdfoutput=1 

\usepackage{jheppub} 
\usepackage[export]{adjustbox}
\usepackage[utf8]{inputenc}
\usepackage{multirow}
\usepackage{bbold}
\usepackage[table]{xcolor}
\usepackage{slashed}
\usepackage{bm}
\usepackage{subfig}
\usepackage{float}
\usepackage{fancyvrb}
\usepackage{fvextra}
\usepackage{soul}
\usepackage[title]{appendix}

\usepackage{physics}
\usepackage{hyperref}
\usepackage{bbm}
\usepackage{graphicx}
\usepackage{adjustbox}

\usepackage{subfig}
\usepackage{color,graphicx,slashed,hyperref}
\usepackage[utf8]{inputenc}
\hypersetup{
   unicode=true,          
   pdftoolbar=true,        
   pdfmenubar=true,        
   pdffitwindow=false,     
   pdfstartview={FitH},    
   pdftitle={SU2FlavourGW},    
   pdfauthor={Author},     
   pdfsubject={Subject},   
   pdfcreator={Creator},   
   pdfproducer={Producer}, 
   pdfkeywords={keyword1} {key2} {key3}, 
   pdfnewwindow=true,      
   colorlinks=true,       
   linkcolor=blue,          
   citecolor=purple,        
   filecolor=purple,      
   urlcolor=blue           
}

\usepackage{comment}
\usepackage{cancel} 
\usepackage{csquotes} 
\usepackage{setspace}

\usepackage{booktabs}
\usepackage{amstext} 
\usepackage{array}   
\newcolumntype{C}{>{$}c<{$}} 
\newcolumntype{L}{>{$}l<{$}} 

\title{Electroweak precision tests for asymptotic Grand Unification models}

\author[a,b]{Giacomo Cacciapaglia,}
\emailAdd{cacciapa@lpthe.jussieu.fr}
\author[c,d]{Aldo Deandrea,}
\emailAdd{deandrea@ip2i.in2p3.fr}
\author[c]{Christian Verollet}
\emailAdd{c.verollet@ip2i.in2p3.fr}

\affiliation[a]{Laboratoire de Physique Th\'eorique et Hautes \'Energies - LPTHE, Sorbonne Universit\'e, CNRS, 4 Place Jussieu, 75005 Paris, France}
\affiliation[b]{Quantum Theory Center ($\hslash$QTC) at IMADA \& D-IAS, Southern Denmark Univ., Campusvej 55, 5230 Odense M, Denmark}
\affiliation[c]{Universit\'e Lyon 1, CNRS, IP2I, UMR 5822, Villeurbanne, France}
\affiliation[d]{Department of Physics, University of Johannesburg, 
PO Box 524, Auckland Park 2006,\\ South Africa}

\abstract{Asymptotic grand unification is an alternative framework to traditional quantitative unification, as the renormalisation flow leads towards an ultra-violet safe fixed point. Phenomenologically, 5-dimensional realisations permit new particles with masses as low as the TeV scale, well below the usual unification scale. We explore the impact of such models on electroweak precision observables, focusing on a minimal SU(5) template  for concreteness. We show that current precision electroweak measurements are not sensitive to this class of models, but bounds in the TeV range can be obtained by recasting LHC analyses. Future colliders, such as CEPC and FCC-ee, can push the 95\% limit on the Kaluza-Klein mass beyond the direct reach of the LHC programme.}


\begin{document} 
\maketitle


\section{Introduction}
Unification models aim at reducing the number of independent couplings and field multiplets compared to lower energy models, such as the Standard Model (SM). The latter is based on the gauge principle and it employs three independent gauge couplings and several matter multiplets to describe quarks, leptons and their interactions. In Grand Unified Theories (GUTs) \cite{Georgi:1974sy,Georgi:1974yf,Ross:1985ai,Mohapatra:1986uf}, it is typically assumed that gauge coupling unification occurs at a specific scale where low-energy couplings converge via their renormalisation group running. 

An alternative is offered by asymptotic unification \cite{Bajc:2016efj}, where couplings tend to the same UV fixed point and, therefore, are only equal at  \emph{asymptotically} high energies. The existence of UV fixed points in gauge-Yukawa theories (without gravity) was established in \cite{Litim:2014uca} and applied to the SM in \cite{Abel:2017rwl,Abel:2018fls}.
A concrete realisation of asymptotic GUT relies on gauge theories in five dimensions (5D) \cite{Dienes:2002bg}, where the gauge symmetry is broken by boundary conditions on an orbifold \cite{Kawamura:1999nj}. This leads to a class of models called \emph{aGUT} (``a'' for asymptotic) \cite{Cacciapaglia:2020qky,Cacciapaglia:2023kyz,Cacciapaglia:2024duu,Cacciapaglia:2025bxs}. Various promising models were studied, based on $SU(5)$ \cite{Cacciapaglia:2020qky,Cacciapaglia:2022nwt}, $SO(10)$ \cite{Khojali:2022gcq,Fang:2025hlc}, $SU(6)$ \cite{Cacciapaglia:2023kyz} and $E_6$ \cite{Cacciapaglia:2023ghp}. One common feature of aGUTs is that SM fermions cannot be embedded within the same representations as in standard GUT, with the consequence of preventing proton decay and allowing for extra dimension scales as low as the TeV scale. In this work we explore the consequences on the electroweak precision observables of aGUT models: for simplicity and concreteness, we focus on the simplest $SU(5)$ aGUT \cite{Cacciapaglia:2020qky}, which allows us to illustrate the general features of the corrections. Nevertheless, the results we obtained can be easily extended to other symmetry breaking patterns. The $SU(5)$ model is therefore a template for more general aGUT models and is chosen due to its minimality and historical significance. As mentioned above, achieving asymptotic unification requires a different arrangement of fermion multiplets due to the presence of a compact extra dimension and the corresponding orbifold parities. This results in a new class of particles, called Indalo-particles, which have unique properties \cite{Cacciapaglia:2020qky}. The Indalo-particles, in fact, have exotic baryon and lepton numbers, preventing their decay into SM particles and thus forbidding proton decay. 
The model also introduces a scalar multiplet containing the Higgs doublet and its Indalo-partners. 

One generic way of probing the effect of heavy new physics above the electroweak scale consists in looking at their effects on vacuum-polarisation diagrams of electroweak bosons, encased in the so-called oblique corrections. These effects have been parameterised by Peskin and Takeuchi \cite{peskin1990,peskin1992} and encapsulated in the three oblique parameters: $S,T,U$ (an equivalent formulation is due to Altarelli and Barbieri \cite{Altarelli:1990zd}). Under a standard set of assumptions, corrections to electroweak observables can be expressed in terms of these three. Specifically, $T$ quantifies the strength of weak isospin breaking through correction to the $W$ and $Z$ masses, while $S$ is an isospin-symmetric measure of the size of the Higgs sector. These parameters are very sensitive to $Z$-pole measurements, and were constrained at LEP. At higher energies, more parameters can be defined \cite{Barbieri:2004qk,Cacciapaglia:2006pk}, corresponding to four-fermion operators. Also, an effective field theory analysis shows that $U$ corresponds to a higher order operator as compared to $S$ and $T$, hence it is often neglected.
The effects on precision electroweak physics from extensions of the SM provide valuable information and restrictions on the possible particle content and structures of new physics models. In particular, future colliders such as the FCC-ee \cite{FCC:2025lpp} and CEPC \cite{CEPCStudyGroup:2023quu} will enable more precise measurements of these parameters \cite{fan_possible_2015} and thus better constraints on the lower bound of heavy new physics. The main contributions to oblique parameters coming from a 5D formulation of the SM  were explored in \cite{Appelquist_2001}; here we propose a detailed calculation in our specific setup, which also extends the general results present in the literature.

The manuscript is organised as follows: in Sec.~\ref{sec:model} we recap the main features of the model, introducing the ghost sector in detail; in Sec.~\ref{sec:ewpts} we present the results for the oblique parameters in the model and the corresponding bounds, before presenting our conclusions in Sec.~\ref{sec:concl}. The appendices contain further details on the computation, providing results that can be easily applied to other 5D models, including aGUT models and beyond.

\section{The model} \label{sec:model}
We consider an aGUT model based on 5D bulk gauge symmetry $SU(5)$, with a breaking pattern leading to the SM group $SU(3) \times SU(2) \times U(1)$. This is achieved via a single extra-dimension, compactified on the orbifold $S^1/(\mathbb{Z}_2\times \mathbb{Z}_2')$ with radius $R$. For a complete description of the model, we refer to \cite{Cacciapaglia:2020qky,Cacciapaglia:2022nwt}.

The particle content of this model is a generalisation of the standard $SU(5)$ unification model containing additional particles (Indalo fields) with masses proportional to $1/R$. They have the same charges of SM fermions and $SU(5)$ GUT extra bosons, but atypical baryon and lepton numbers. Hence, we will denote them by a capital letter corresponding to the lower-case one of the matching SM field (plus $X$, $Y$ and $H$ for the additional gauge bosons and coloured Higgs, as in standard $SU(5)$ \cite{Georgi:1974sy}). 
The fermion sector differs from the usual GUTs, as SM fermions cannot all be embedded into a single set of $SU(5)$ representations. Instead, $SU(5)$ aGUT consists of a multiplet in the $\overline{\textbf{5}}$-representation that contains the SM lepton doublet and the Indalo bottom singlet partner, as follows:
\begin{equation}
    \psi_{\overline{5}_{L/R}} = 
\begin{pmatrix} 
    B^c \\ l
\end{pmatrix}_{L/R}
=
\begin{pmatrix} 
B^{c(1)} \\B^{c(2)} \\ B^{c(3)} \\ l_{\tau^-} \\ -l_{\overline{\nu}_{\tau}}
\end{pmatrix}_{L/R}\,.
\end{equation}
This multiplet contains the zero-modes for the left-handed lepton doublet $l_L$, together with a new heavy colour-triplet field $B^c$. One additional $\textbf{5}$-representation is required to contain the $b$-quarks together with a new Indalo lepton doublet $L^c$, as follows:
\begin{equation}
    \psi_{5_{L/R}} = \begin{pmatrix} b \\ L^c
\end{pmatrix}_{L/R}=\begin{pmatrix} b^{(1)} \\b^{(2)} \\ b^{(3)} \\ L_{\mathcal{T}}^c \\ -L_{\mathcal{N}}^c
\end{pmatrix}_{L/R}\,.
\end{equation}
This doubling of the standard 5-plet of $SU(5)$ is necessary in this model due to parity and quantum number considerations, because of the extra-dimensional nature of the model. We add a subscript $L$ or $R$ to keep track of the chirality of the fermion components (we recall that a 5D fermion field contains both 4D chiralities). 
Similarly, the standard 10-plet structure must be doubled, via a \textbf{10}-representation containing the SM quark doublet zero-mode $q$ together with an Indalo top singlet $T^c$ and Indalo tau singlet $\mathcal{T}^c$, as follows:
\begin{equation}
\psi_{10_{L/R}} = \frac{1}{\sqrt{2}}\begin{pmatrix} T^c & q \\ & \mathcal{T}^c
\end{pmatrix}_{L/R} = \frac{1}{\sqrt{2}}\begin{pmatrix} 0 & T^{c(3)} & - T^{c(2)} & q_t^{(1)} & q_b^{(1)} 
\\ -T^{c(3)} & 0 & T^{c(1)} & q_t^{(2)} & q_b^{(2)}
\\ T^{c(2)} & - T^{c(1)} & 0 & q_t^{(3)} & q_b^{(3)}
\\ -q_t^{(1)} & -q_t^{(2)} & -q_t^{(3)} & 0 & \mathcal{T}^c
\\ -q_b^{(1)} & -q_b^{(2)} & -q_b^{(3)} & - \mathcal{T}^c & 0
\end{pmatrix}_{L/R}\,,
\end{equation}
and another $\overline{\textbf{10}}$-representation containing the zero-modes of lepton and top-quark singlets, $\tau$ and $t$, and the Indalo quark doublet $Q^c$. as follows:

\begin{equation}
\psi_{\overline{10}_{L/R}} = \frac{1}{\sqrt{2}}\begin{pmatrix} t & Q^c \\ & \tau
\end{pmatrix}_{L/R} = \frac{1}{\sqrt{2}}\begin{pmatrix} 0 & t^{(3)} & - t^{(2)} & Q_{T}^{c(1)} & Q_{B}^{c(1)} 
\\ -t^{(3)} & 0 & t^{(1)} & Q_{T}^{c(2)} & Q_{B}^{c(2)}
\\ t^{(2)} & - t^{(1)} & 0 & Q_{T}^{c(3)} & Q_{B}^{c(3)}
\\ -Q_{T}^{c(1)} & -Q_{T}^{c(2)} & -Q_{T}^{c(3)} & 0 & \tau
\\ -Q_{B}^{c(1)} & -Q_{B}^{c(2)} & -Q_{B}^{c(3)} & - \tau & 0
\end{pmatrix}_{L/R}\,.
\end{equation}
Right-handed neutrinos can be introduced via singlet fields, see for instance \cite{Alezraa:2025lct}, we will not include them here as they do not contribute significantly to oblique parameters.

The SM Higgs doublet $\phi_h$ is contained in a scalar $\bf{5}$-plet, as in the standard $SU(5)$ model:
\begin{equation}
    \phi_5 = \begin{pmatrix} H \\ \phi_h 
\end{pmatrix} = \begin{pmatrix} H^{(1)} \\H^{(2)} \\ H^{(3)} \\ \phi^+ \\ -\phi_0
\end{pmatrix}\,,
\end{equation}
however, the colour-triplet Higgs $H$ has Indalo nature, so its mass is naturally split from that of the doublet Higgs zero mode \cite{Kawamura:2000ev}.
Two parities $P_0$ and $P_1$ act on the two fixed points $y = 0$ and $y = \pi R / 2$ of the extra space dimension.
Each 5D field can be decomposed into towers of Kaluza-Klein (KK) modes, corresponding to 4D particle fields, whose characteristics depend on the parity under the $\mathbb{Z}_2$ and the $\mathbb{Z}'_2$ symmetries, which we denote as $(\pm, \pm)$ and $(\pm, \mp)$. The most important characteristic is the presence or absence of zero-modes that can be identified as SM particles; instead, higher modes have masses of the order $1/R$. Zero-modes appear only when the parity signs for the field components are $(+,+)$. The parities of the $L$ and $R$ chiralities of the same 5D field are opposed, hence ensuring that fermion zero-modes are chiral. More details on the decomposition can be found in \cite{Cacciapaglia:2020qky,Cacciapaglia:2022nwt}. We recall that all modes will have KK masses $n/R$, with $n$-odd for Indalo fields with parities $(\pm,\mp)$, and $n$-even for all fields with parities $(\pm,\pm)$, including the SM ones. The model is described by a local 5D gauge-invariant Lagrangian density, invariant under $SU(5)$:
\begin{equation}
\begin{aligned}
 &{\cal{L}}_{\mathrm{SU5}}(x,y) = \\
\text{(Gauge bosons sector)} \quad &-\frac{1}{4}F^a_{MN}{F^a}^{{MN}} \\
\text{(Gauge fixing part)}\quad &-\frac{1}{2\xi}\left(\partial_{\mu}{A}^{a\mu}-\xi\left(\partial_{y}A^a_5+g_2{F'}^a\chi'_5\right)\right)^2\\
\text{(Fadeev-Popov ghosts)}\quad &-\bar{\eta}^a\left(-\partial^{\mu}D^{ab}_{\mu}+\xi\left[\partial_{y}D^{ab}_y-(F'F'^T)^{ab}-g_2(F'(T'\chi_5')^T)^{ab}\right]\right)\eta^b\\
\text{(Fermionic sector)}\quad&+ \bar{\psi}_{5}i\slashed{D}\psi_{5}+\bar{\psi}_{\bar{5}}i\slashed{D}\psi_{\bar{5}}+\mathrm{Tr}\left(\bar{\psi}_{10}i\slashed{D}\psi_{10}\right)+\mathrm{Tr}\left(\bar{\psi}_{\bar{10}}i\slashed{D}\psi_{\bar{10}}\right)\\
\text{(Yukawa sector)}\quad&-\sqrt{2}\left(Y_{\tau}\bar{\psi}_{\bar{5}}\psi_{\bar{10}}\phi^{*}_{5}+Y_{b}\bar{\psi}_{5}\psi_{10}\phi^{*}_{5}+\frac{1}{2}Y_{t}\epsilon_{5}\bar{\psi}_{\bar{10}}\psi_{10}\phi_{5}+\mathrm{h.c}\right)\\
\text{(Higgs sector)}\quad&+\abs{D_{M}\phi_{5}}^2-V\left(\phi_5\right)\,.
\end{aligned}
\label{sect2.eq1}
\end{equation}
For the purpose of the calculations, the gauge fixing part is given explicitly: Fadeev-Popov ghosts are denoted as $\eta^a$ and $\bar{\eta}^a$, since in non-unitary gauge their loop contributions are essential in order to have a UV finite result.  Following the standard procedure \cite{PeskinSchroeder}, a supplementary matrix $F$ and a vector $\chi_5$ are needed to describe, in a compact way, the gauge fixing part of the theory and the Fadeev-Popov ghosts. The notations and definitions used here are detailed in \hyperref[App A]{Appendix A}. 

To perform the calculation in the usual 4D quantum field theory framework, we dimensionally reduce the 5D action to an effective 4D one by performing the integration over the compactified dimension 
\begin{equation}
S = \int\dd[4]{x}\dd{y}{\cal{L}}_{\mathrm{SU5}}(x,y) = \int\dd[4]{x}{\cal{L}}_{eff}\,.
\label{sect2.eq2}
\end{equation}
This has the well-known effect of giving a mass $M_n = n/R$ to each particle of the KK tower, according to their parity. This property can be understood as the conservation of the fifth momentum component along the compactified dimension, in a discretised version. 

In addition to the KK masses, particles that couple to the scalar doublet also get a mass from the usual Higgs mechanism. The SM zero modes receive their usual masses, while the supplementary Indalo gauge bosons and the coloured scalar triplet receive a contribution $\sim m_W^2$. For the fermions, the Higgs sector mixes the different chiralities of the doublet and singlet components. For the purpose of our computation of oblique corrections, we will be interested only in the top and bottom quarks and their Indalo partners. The reason is that KK states are vector-like, hence they do not contribute to electroweak observables unless their masses are split by the Higgs, hence their contribution to oblique parameters will be proportional to their Higgs-induced mass. Of all fermions, only the top has a sizeable mass, $m_t = 172.57$ GeV \cite{ParticleDataGroup:2024cfk}, hence the top and the Indalo-top towers will give the dominant contribution.
The mass matrix for the $n$-mode top doublet and singlet has the following form
\begin{equation}
-{\begin{pmatrix}
\overline{t}\:\:\:\: & \overline{q}_{t}
\end{pmatrix}}
\begin{pmatrix}
-M_{n} &  m_{t}\\
 m_{t} & M_{n}\\
\end{pmatrix}
\begin{pmatrix}
t\\
q_t
\end{pmatrix}\qq{with} M_n = \frac{n}{R}\;,\
\label{sub2.1:eq1}
\end{equation}
where a similar matrix describes the $b,q_b,T,B,Q_T,Q_B$ components. The mass eigenvalues are degenerate
\begin{equation}
M_{n}^{t}= \sqrt{M_{n}^2+m_{t}^2}\,,
\label{sub2.1:eq2}
\end{equation}
while the mass eigenstates, denoted by primed fields, are defined as \cite {Appelquist_2001}
\begin{equation}
\begin{pmatrix}
t \\ q_t 
\end{pmatrix}=\begin{pmatrix}
    -\gamma_5\cos\alpha^t_n & \sin\alpha^t_n \\
    \gamma_5\sin\alpha^t_n & \cos \alpha^t_n
\end{pmatrix}
\begin{pmatrix}
    t'\\q_t'
\end{pmatrix}\qq{with}\tan2\alpha^t_n = \frac{m_t}{M_n}\,.
\label{part2:eq10}
\end{equation}
A similar mixing occurs for the Goldstone bosons from the Higgs multiplet $\phi_5$ with the gauge scalars $Z_5,W_5^{\pm}$ and $Y_5$, see Eq.(\ref{AppA:eq4}) in \hyperref[App A]{Appendix A}. The main difference is that the mass matrices depend on the gauge-fixing parameter $\xi$ and, by choosing the 't Hooft-Feynman gauge i.e $\xi=1$, the matrices are diagonal and each of these scalars ends up with a mass equal to that of the corresponding gauge boson mode $M^V_{n}$, where $V = W$ for $Y_5,W^{\pm}$ and $Z$ for $Z_5$.

\section{Electroweak precision observables and bounds} \label{sec:ewpts}

\begin{figure}[ht!]
    \centering
    \includegraphics[scale=1.2]{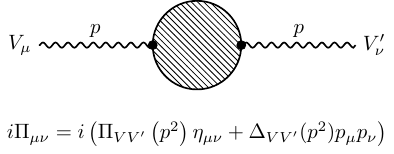}
    \caption{Self-energy for a $V_{\mu}$ and $V'_{\nu}$ zero-mode electroweak vector, the $VV'$ combinations possible and used in the calculations are $W^+W^+,W^-W^-,ZZ,\gamma\gamma$ and $Z\gamma$.}
    \label{fig1}
\end{figure}

The oblique parameters $S,T,U$  were originally defined in \cite{peskin1992} as
\begin{equation}
\begin{aligned}
&\alpha \,S = 4e^2\left[{\Pi^{new}_{33}}'\left(0\right)-{\Pi^{new}_{3Q}}'\left(0\right)\right]\,,\\
&\alpha\, T = \frac{e^2}{s^2_wc^2_wm_Z^2}\left[{\Pi^{new}_{11}}\left(0\right)-{\Pi^{new}_{33}}\left(0\right)\right]\,,\\
&\alpha\, U = 4e^2\left[{\Pi^{new}_{11}}'\left(0\right)-{\Pi^{new}_{33}}'\left(0\right)\right]\,,
\label{sub3.1:eq1}
\end{aligned}
\end{equation}
in terms of the vacuum polarisation functions $\Pi$ in the gauge basis.
The label `$new$' indicates that only non-SM particles are involved in the loops of the self-energies, 
as these parameters are defined up to SM loop corrections (in contrast, the equivalent formulation in terms of Altarelli-Barbieri parameters \cite{Altarelli:1990zd} contains the SM corrections). These definitions involve only the $\eta_{\mu\nu}$ part of the self-energies, see Figure \ref{fig1}. As pointed out by the authors of \cite{peskin1992}, this simplification is due to the fact that they consider only processes involving light fermions as external particles, since those are most readily available for high precision measurements at present-day experiments. As a consequence, the terms proportional to $p_{\mu}p_{\nu}$ in the $W$ and $Z$ propagators can be neglected.~\footnote{This is because contraction with external fermion currents suppresses the $p_{\mu}p_{\nu}$ terms compared to the $\eta_{\mu\nu}$ terms by a factor of $m _f^2 /m_Z^2$ where $m_f$ is the external fermion mass.}

One very important feature of the oblique parameters is that they are defined such that the subtraction of the combination of self-energies removes the UV divergences from the loops. The oblique parameters in Eq.(\ref{sub3.1:eq1}) are expressed in the gauge basis ($11,33,3Q$), which is related to the mass basis as
\begin{equation}
\begin{aligned}
&\Pi_{11}(p^2)=\frac{1}{2g^2}\left[\Pi_{W^+W^+}(p^2)+\Pi_{W^-W^-}(p^2)\right]\;,\\
&\Pi_{3Q}(p^2) = \frac{1}{g^2}\left[\Pi_{\gamma\gamma}(p^2)+\frac{c_w}{s_w}\Pi_{Z\gamma}(p^2)\right]\;,\\
&\Pi_{33}(p^2) =\frac{1}{g^2}\left[c_w^2\Pi_{ZZ}(p^2)+2s_wc_w\Pi_{Z\gamma}+s^2_w\Pi_{\gamma\gamma}(p^2)\right] \,.
\label{sub3.1:eq3}
\end{aligned}
\end{equation}

\subsection{Fermionic loop contributions}
\begin{figure}[ht!]
    \centering
    \includegraphics[scale=1.2]{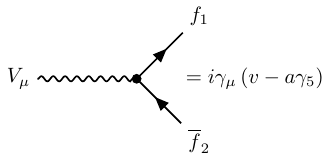}\\
    \caption{Generic Feynman rule for the interaction vertex of a zero-mode electroweak vector $V_{\mu}$ and two non-zero modes fermions $f_1$ and $f_2$. When the fermions are charge conjugate, the vertex rule change to $-i\gamma_{\mu}^T\left(v-a\gamma_5^T\right)$.}
    \label{fig2}
\end{figure}

\begin{figure}[ht!]
    \centering
    \includegraphics[scale=1.2]{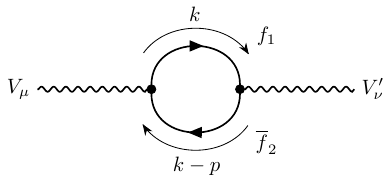}
    \caption{Diagram for the self-energy of zero-mode electroweak vectors, $V_{\mu}$ into $V'_{\nu}$, with two fermions $f_i$, with $i=1,2$ and mass $m_1,m_2$, circulating in the loop. The associated amplitude is given in Eq.(\ref{sub3.2:eq1}).}
    \label{fig3}
\end{figure}

The contribution of fermionic loops is usually the most important, because of the explicit breaking of the custodial symmetry by the top mass. All the interaction vertices of fermions with electroweak vectors can be written in a generic way, see  Fig. \ref{fig2}, with a vector coupling $v$ and an axial one $a$.
\\It follows that all loops associated to a couple of fermions, illustrated in Fig. \ref{fig3}, can be written using the above Feynman rule as
\begin{equation}
\begin{aligned}
i\Pi_{\mu\nu}(p^2)=- &\int\frac{\mathrm{d}^4k}{\left(2\pi\right)^4}\mathrm{Tr}\left[i\gamma_{\mu}\left(v-a\gamma_5\right)\dfrac{i\left(\slashed{k}+m_1\right)}{k^2-m_1^2}i\gamma_{\nu}\left(v'-a'\gamma_5\right)\dfrac{i\left(\slashed{k}-\slashed{p}+m_2\right)}{\left(k-p\right)^2-m_2^2}\right]\,.
\end{aligned}
\label{sub3.2:eq1}
\end{equation}
For charge conjugate fermions, the charge operator changes the matrices to their transpose with a minus sign, but this does not change the result because of the trace of the gamma matrices in the total amplitude.
\\We express the result in terms of Passarino-Veltman (PV) functions:
\begin{equation}
\begin{aligned}
\Pi_{VV'}(p^2) =&\frac{2}{16\pi^2}\biggl\{(aa'+vv')\left[A_0(m_1)+A_0(m_2)-4B_{00}(p^2,m_1,m_2)\right]\\&+\left[aa'\left((m_1+m_2)^2-p^2\right)+vv'\left((m_1-m_2)^2-p^2\right)\right]B_0(p^2,m_1,m_2)\biggr\},
\end{aligned}
\end{equation}
where we provide the detailed convention and computation in \hyperref[App C]{Appendix C}. A crucial feature is that the contribution of the PV functions depends on the mass splitting $m_1^2-m_2^2$ between the two fermions in the loop, and it is minimal when $m_1 = m_2$. Because of this, we can limit our computation to the quark loops, where the custodial symmetry is maximally broken by the top mass. It is the tower of top KK modes and a tower of Indalo states, which also received a contribution to the mass via the top Yukawa, that contributes to the loop. When we neglect the natural mass of the bottom mass quark, which is different between the SM-like states $b,q_b$ and the Indalo ones $B,Q_B$, both towers give the same contribution. However, different KK masses $M_n = n/R$ enter the result: for the top tower, $n$ is even, while for the Indalo tower $n$ is odd. Using the Feynman rules shown in \hyperref[App A]{Appendix A}, where we neglect the natural mass of the bottom quark, the loops in Fig. \ref{fermionic_loops_WW} and \ref{fermionic_loops_33} give
\begin{equation}
\begin{aligned}
\Pi^n_{11}(p^2) &= \frac{2}{16\pi^2}\frac{1}{4}\biggl\{2A_0(M_n^t)+2A_0(M_n)-4B_{00}(p^2,M_n^t,M_n)-4B_{00}(p^2,M_n,M_n^t)\\&\quad\quad\quad\quad\quad\quad
+\left[m_t^2-p^2\right]\left(B_0(p^2,M_n^t,M_n)+B_0(p^2,M_n,M_n^t)\right)\biggr\},\\
\Pi_{33}^n(p^2) &=  \frac{2}{16\pi^2}\frac{1}{4}\biggl\{2A_0(M_n^t)-4B_{00}(p^2,M_n^t,M_n^t)+\left[2m_t^2-p^2\right]B_0(p^2,M_n^t,M_n^t)\\&+2A_0(M_n)-p^2B_0(p^2,M_n,M_n)-4B_{00}(p^2,M_n,M_n)\biggr\},\\
\Pi^n_{3Q}(p^2) &=\frac{2}{16\pi^2}\frac{1}{4}\biggl\{2Q_t\left[2A_0(M_n^t)-4B_{00}(p^2,M_n^t,M_n^t)-p^2B_0(p^2,M_n^t,M_n^t)\right]\\&\quad\quad\quad\quad\quad-2Q_b\left[2A_0(M_n)-4B_{00}(p^2,M_n,M_n)-p^2B_0(p^2,M_n,M_n)\right]\biggr\}\,,
\end{aligned}
\end{equation}
where $M_n^t$ is defined in Eq.~\eqref{sub2.1:eq2}.
Once we write each PV function in terms of the divergent and scale-dependent part $\Delta = 1/\bar{\epsilon}-\ln(M_n^2/\mu^2)$ and the small parameter $z = m_t^2/M_n^2 \ll 1$ -- see \hyperref[App B]{Appendix B}, we show that the oblique parameters contribution from a single mode $n\in \mathbbm{N}$ can be written as
\begin{equation}
\begin{aligned}
\alpha\, S^n =& \frac{3e^2}{8\pi^2}\biggl\{\frac{z}{3(1+z)}-\frac{2}{9}\ln(1+z)\biggr\}\:,\\
\alpha\,T^n=&\frac{3m_t^2}{16\pi^2v^2}\biggl\{1-\frac{2}{z}+\frac{2}{z^2}\ln(1+z)\biggr\}\:,\\
\alpha\, U^n =& \frac{3e^2}{8\pi^2} \biggl\{\frac{12+42z+16z^2-17z^3}{9 z^2 (z+1)} -\frac{4(1+4z+3z^2-z^3-z^4)}{3z^3 (z+1)}\ln(1+z)\biggr\}\:,
\end{aligned}
\end{equation}
where, as predicted, all $\Delta$ terms cancel. Then, expanding in $z$ and taking the sum on all odd and even indices, the main contribution to oblique parameters from the top sector is given by
\begin{equation}
\begin{aligned}
&\alpha\,S^t = \frac{3\alpha}{2\pi}\sum_{n\in\mathbbm{N}}\left\{\frac{m_t^2}{9M_n^2}+\dots\right\} =  \frac{\alpha\pi m_t^2R^2}{36}+\dots\,, \\
&\alpha\,T^t = \frac{3m_t^2}{16\pi^2 v^2}\sum_{n\in\mathbbm{N}}\left\{ \frac{2m_t^2}{3M_n^2}+\dots\right\} = \frac{m_t^4R^2}{48v^2}+ \dots\,,\\
&\alpha\,U^t = \frac{3\alpha}{2\pi}\sum_{n\in\mathbbm{N}} \left\{\frac{2m_t^4}{5M_n^4}+\dots\right\} =  \frac{\alpha\pi^3m_t^4R^4}{75}+\dots\,. 
\end{aligned}
\label{sub3.2:eq4}
\end{equation}
where the dots contain higher order terms, with $\alpha = e^2/4\pi$ being the fine structure constant at the electroweak scale. We used the following sums, running over all non-zero positive integers:
\begin{equation}
\sum_{n\in\mathbbm{N}} \frac{1}{M_n^2} = R^2 \sum_{n\in\mathbbm{N}} \frac{1}{n^2} = \frac{\pi^2 R^2}{6}\;,\qquad
\sum_{n\in\mathbbm{N}} \frac{1}{M_n^4} = R^4 \sum_{n\in\mathbbm{N}} \frac{1}{n^4} = \frac{\pi^4 R^4}{90}\;.
\end{equation}
It is important to highlight the expected fact that $U$ received the first non-vanishing contribution at higher order than $S$ and $T$. Also, we emphasised here that only in the configuration where the different natural masses of the bottom quark and its Indalo partner are neglected, the two different Kaluza-Klein towers merge together and give the same result for $T$ as in \cite{Appelquist_2001}.

\subsection{Scalar and vector loop contributions} 

The number of one-loop diagrams from this sector is large, hence a very helpful trick consists in separating the odd (Indalo) and even KK modes. Hence, the set of all loops can be separated into three independent sets: the first one consists of all the massive KK modes of the SM zero modes, corresponding to even modes; the second contains all the new even modes not present in the SM (essentially, the gauge scalars); finally the third consists of all odd (Indalo) modes. The different Feynman rules of the vertices involved in the loops, computed with Feynrules \cite{FeynRule},  can be found in \hyperref[App E]{Appendix E} in addition to the different propagators in the $R_{\xi}$ gauge. 

For the first set, consisting of the SM-like loops, we can adapt calculations for the SM \cite{BardinBible} to our case. A summary of all loops and associated PV functions can be seen in Fig. \ref{fig5}, \ref{fig6}  and in Eq.(\ref{AppC:eq4}), (\ref{AppC:eq5}). Inside the SM-like set, we can identify two independent subset where the cancellation of the $\Delta$--terms occur. The first one, involving the Higgs and the associated self-energies, read  
\begin{equation}
\begin{aligned}
&\Pi_{11}^n(p^2) = \frac{1}{16\pi^2}\biggl\{-m_W^2B_0\left(p^2,M_n^h,M_n^W\right)+\frac{1}{4}A_0\left(M_n^h\right)-\frac{s_w^4}{c_w^2}m_W^2B_0\left(p^2,M_n^Z,M_n^W\right)\\&\quad\quad\quad\quad\quad-s^2_wm^2_WB_0\left(p^2,M_n,M_n^W\right)+B_{00}\left(p^2,M_n^W,M_n^h\right)+B_{00}\left(p^2,M_n^Z,M_n^W\right)\biggr\}\;,\\
&\Pi_{33}^n(p^2) = \frac{1}{16\pi^2}\biggl\{-m_Z^2B_0\left(p^2,M_n^h,M^Z_n\right)+\frac{1}{4}A_0\left(M_n^h\right)+B_{00}\left(p^2,M_n^Z,M_n^h\right)\\&
\quad\quad\quad\quad\quad\quad\quad\quad+B_{00}\left(p^2,M_n^W,M_n^W\right)\biggr\}\;,\\&\Pi_{3Q}(p^2) = \frac{1}{16\pi^2}\biggl\{2B_{00}\left(p^2,M_n^W,M_n^W\right)\biggr\}\;.
\end{aligned}
\label{sub3.3:eq1}
\end{equation}
It follows that all three parameters receive a contribution involving the Higgs mass:

\begin{equation}
\begin{aligned}
&\alpha\,S^{SM-like}_h =\frac{\alpha}{\pi}\sum_{n=\text{even}} \left\{\frac{m_h^2-(3+2c_w^2)m_Z^2}{24M_n^2}+\dots \right\}= \frac{\alpha\pi R^2(m_h^2-(3+2c^2_w)m_Z^2)}{576}+\dots\,,\\
&\alpha\,T^{SM-like}_h = -\frac{\alpha}{4\pi c_w^2}\sum_{n = \text{even}}\left\{\frac{5m_h^2+7m_W^2}{12M_n^2}+\dots\right\} = -\frac{\alpha\pi  R^2(5m_h^2+7m_W^2)}{1152c_w^2}+\dots\,,\\
&\alpha\,U^{SM-like}_h = -\frac{\alpha}{\pi}\sum_{n=\text{even}} \left\{\frac{s_w^2m_Z^2(5m_h^2+6m_W^2)}{15M_n^4}+\dots \right\}\\&\quad\quad\quad\quad\quad= -                                                \frac{\alpha\pi^3R^4 s_w^2m_Z^2(5m_h^2+4m_W^2)}{21600}+\dots \,,\\
\end{aligned}
\label{sub3.3:eq2}
\end{equation}
where $n$ is an even index since all the particles involved have a zero-mode, leading to the sums
\begin{equation}
\sum_{n=\text{even}} \frac{1}{M_n^2} = \frac{\pi^2 R^2}{24}\;, \qquad \sum_{n=\text{even}} \frac{1}{M_n^4} = \frac{\pi^4 R^4}{1440}\;. 
\end{equation}
The second subset of loops contains only vectors and Goldstone bosons. The self-energies read
\begin{equation}
\begin{aligned}
\Pi_{11}^n(p^2)=\frac{1}{16\pi^2}\biggl\{&c^2_wA_1^1\left(p^2,M_n^Z,M_n^W\right)+s^2_wA_1^1\left(p^2,M_n,M_n^W\right)+\frac{1}{4}A_0\left(M_n^Z\right)\\&+A_3^1\left(M_n^W\right)+c_w^2A_3^1\left(M_n^Z\right)+s^2_wA_3^1\left(M_n\right)+\frac{1}{2}A_0\left(M_n^W\right)\\&-2c^2_wB_{00}\left(p^2,M_n^Z,M_n^W\right)-2s^2_wB_{00}\left(p^2,M_n,M_n^W\right)\biggr\}\;,\\
\Pi_{33}^n(p^2)=\frac{1}{16\pi^2}\biggl\{&A^1_1\left(p^2,M_n^W,M_n^W\right)+2A_3^1\left(M_n^W\right)+\frac{1}{4}A_0\left(M_n^Z\right)+\frac{1}{2}A_0\left(M_n^W\right)\\&-2B_{00}\left(p^2,M_n^W,M_n^W\right)\biggr\}\;,\\
\Pi_{3Q}^n(p^2)=\frac{1}{16\pi^2}\biggl\{&A^1_1\left(p^2,M_n^W,M_n^W\right)+2A_3^1\left(M_n^W\right)-2B_{00}\left(p^2,M_n^W,M_n^W\right)\biggr\}\;,
\end{aligned}
\end{equation}
and thus do not contribute to $S$. The contribution to $T$ and $U$ are 
\begin{equation}
\begin{aligned}
&\alpha\,T_V^{SM-like} = -\frac{\alpha}{4\pi c_w^2}\sum_{n=\text{even}}\biggl\{\frac{11m_W^2}{6M_n^2}+\dots\biggr\} =-\frac{11\alpha\pi R^2m^2_Z}{576}+\dots\;,\\
&\alpha\,U_V^{SM-like} =-\frac{\alpha}{\pi}\sum_{n=\text{even}}\biggl\{\frac{s_w^2m_Z^2m_W^2}{12M_n^4}+\dots\biggr\} =-\frac{\alpha\pi^3R^4s_w^2m_Z^2m_W^2}{17280}+\dots\;.
\end{aligned}
\end{equation}

For the second set, which contains the new even-mode loops, the contributions of the gauge scalar loops are drawn in Fig. \ref{fig7}, \ref{fig8} and the associated PV functions Eq.(\ref{AppC:eq6}), (\ref{AppC:eq7}). We find
\begin{equation}
\begin{aligned}
\Pi_{11}^n(p^2)&=\frac{1}{16\pi^2}\biggl\{s^2_wB_{00}\left(p^2,M_n,M_n^W\right)+c_w^2B_{00}\left(p^2,M_n^Z,M_n^W\right)\\&\quad\quad\quad\quad\quad+A_0\left(M_n^W\right)+s^2_wA_0(M_n)+c^2_wA_0\left(M_n^Z\right)\biggr\}\;,\\
\Pi_{33}^n(p^2)&= \Pi^n_{3Q}(p^2) = \frac{1}{16\pi^2}\biggl\{B_{00}\left(p^2,M_n^W,M_n^W\right)+2A_0\left(M_n^W\right)\biggr\}\;.
\end{aligned}
\label{sub3.3:eq4}
\end{equation}
As for the gauge bosons subset in the SM-like set, there is no contribution to $S$ and we find: 
\begin{equation}
\begin{aligned}
&\alpha\,T^{new}_{even} = -\frac{\alpha}{4\pi c^2_w}\sum_{n=\text{even}} \left\{\frac{7m_W^2}{12M_n^2}+\dots\right\}=-\frac{7\alpha\pi m_Z^2 R^2}{1152}+\dots\,,\\
&\alpha\,U^{new}_{even} = -\frac{\alpha}{\pi} \sum_{n=\text{even}} \left\{\frac{s^2_wc^2_wm_Z^4}{80M_n^4}+\dots\right\} = -\frac{\alpha\pi^3s^2_wc^2_wm_Z^4R^4}{91200}+\dots\,.
\end{aligned}
\label{sub3.3:eq5}
\end{equation}

Finally, the odd Indalo states also provide a contribution. The calculations are simplified because all the particles have the same KK mass $M^W_n$ with odd $n$. The different loops contributing to the self-energies of the electroweak vector bosons can be seen in Fig.\ref{fig9}, \ref{fig10} and Eq.(\ref{AppC:eq8}), (\ref{AppC:eq9}). In the $11,33,3Q$ basis, they read
\begin{equation}
\begin{aligned}
\Pi_{11}^n(p^2) =& \;\Pi_{33}^n(p^2) = \frac{3}{16\pi^2} \biggl\{\frac{1}{2}A_1^1\left(M_n^W\right)+A_3^1\left(M_n^W\right)-\frac{1}{2}\left(M_n^W\right)^2B_0(p^2,M_n^W,M_n^W)\\&\quad\quad\quad\quad\quad\quad-2B_{00}(p^2,M_n^W,M_n^W)+A_0\left(M_n^W\right)\biggr\}\;,\\
\Pi_{3Q}^n(p^2) =&  \frac{3}{16\pi^2} \biggl\{\frac{1}{2}A_1^1\left(M_n^W\right)+A_3^1\left(M_n^W\right)-\frac{1}{2}\left(M_n\right)^2B_0(p^2,M_n^W,M_n^W)\\&\quad-2Q_YI_Ym_W^2B_0(p^2,M_n^W,M_n^W)-2B_{00}(p^2,M_n^W,M_n^W)+A_0\left(M_n^W\right)\biggr\}\;.
\end{aligned}
\label{sub3.3:eq6}
\end{equation}
As a consequence, the contribution of Indalo particles to $T$ and $U$ are zero, and the one to  $S$ involves only $m_W$ mass and a sum on odd $n$.  We find
\begin{equation}
\begin{aligned}
&\alpha\,S^{new}_{odd} = -\frac{\alpha}{\pi}\sum_{n=\text{odd}} \left\{ \frac{5m_W^2}{36M_n^2}+\dots\right\} = -\frac{5\alpha\pi m_W^2 R^2}{288} +\dots \;.\\
\end{aligned}
\label{sub3.3:eq7}
\end{equation}
where the sum over odd $n$ gives
\begin{equation}
    \sum_{n=\text{odd}} \frac{1}{M_n^2} = \frac{\pi^2 R^2}{8}\,.
\end{equation}
Summing all the contributions from the bosonic sector, we find: 

\begin{equation}
\begin{aligned}
    & \:\alpha S^g = \frac{\alpha\pi R^2(m_h^2-(3+7c^2_w)m_Z^2)}{576} + \dots\,, \\
    &\:\alpha T^g = -\frac{\alpha\pi R^2(5m_h^2+36m_W^2)}{1152c^2_w}+\dots\,, \\
    & \:\alpha U^{g} = {\cal{O}}\left(R^4\right) \;.
\end{aligned}
\label{eq:totalbosons}
\end{equation}
where $U$ arises at higher order, as expected. 


\subsection{Experimental bounds and comparison for future colliders}
In the SM we have by definition $S=T=U=0$, in agreement with the current measured values \cite{ParticleDataGroup:2024cfk}:
\begin{equation}
\begin{aligned}
&S = -0.04\pm0.10\\
&T = 0.01\pm0.12\\
&U =-0.01\pm0.09
\end{aligned}\qq{with correlations}
\begin{tabular}{c|c c c}
 &  S \  &  T & U\\ \hline
  S   &  1 & 0.93 & -0.70\\
  T & & 1 & -0.87 \\
  U & & & 1
\end{tabular}.
\label{STU_SM}
\end{equation}

For future colliders, combining the expected measurements from HL-LHC, projections for errors on $S$ and $T$ have been obtained in \cite{deBlas:2019rxi,FCC2020} for both FCC-ee and CEPC, separately. Assuming a null central value, their $1\sigma$ variations are expected to be
\begin{equation}
   \begin{tabular}{c|c|c}
   & CEPC & FCC-ee  \\ \hline
 S & 0.0068 &0.0038 \\
  T & 0.0072 &0.0022
\end{tabular}
 \label{STU_FCC_CEPC}
\end{equation}
where the correlations can be read off from Fig. 17 of \cite{deBlas:2019rxi}.
To compare the $SU(5)$ aGUT predictions with the data, we sum every contribution to the self energies coming from the fermionic loops in Eq.~\eqref{sub3.2:eq4} and the bosonic loops in Eq.~\eqref{eq:totalbosons}. 

\begin{figure}[ht!]
    \centering
    \begin{adjustbox}{center}
        \begin{minipage}{0.5\textwidth}
            \centering
            \includegraphics[width=\textwidth]{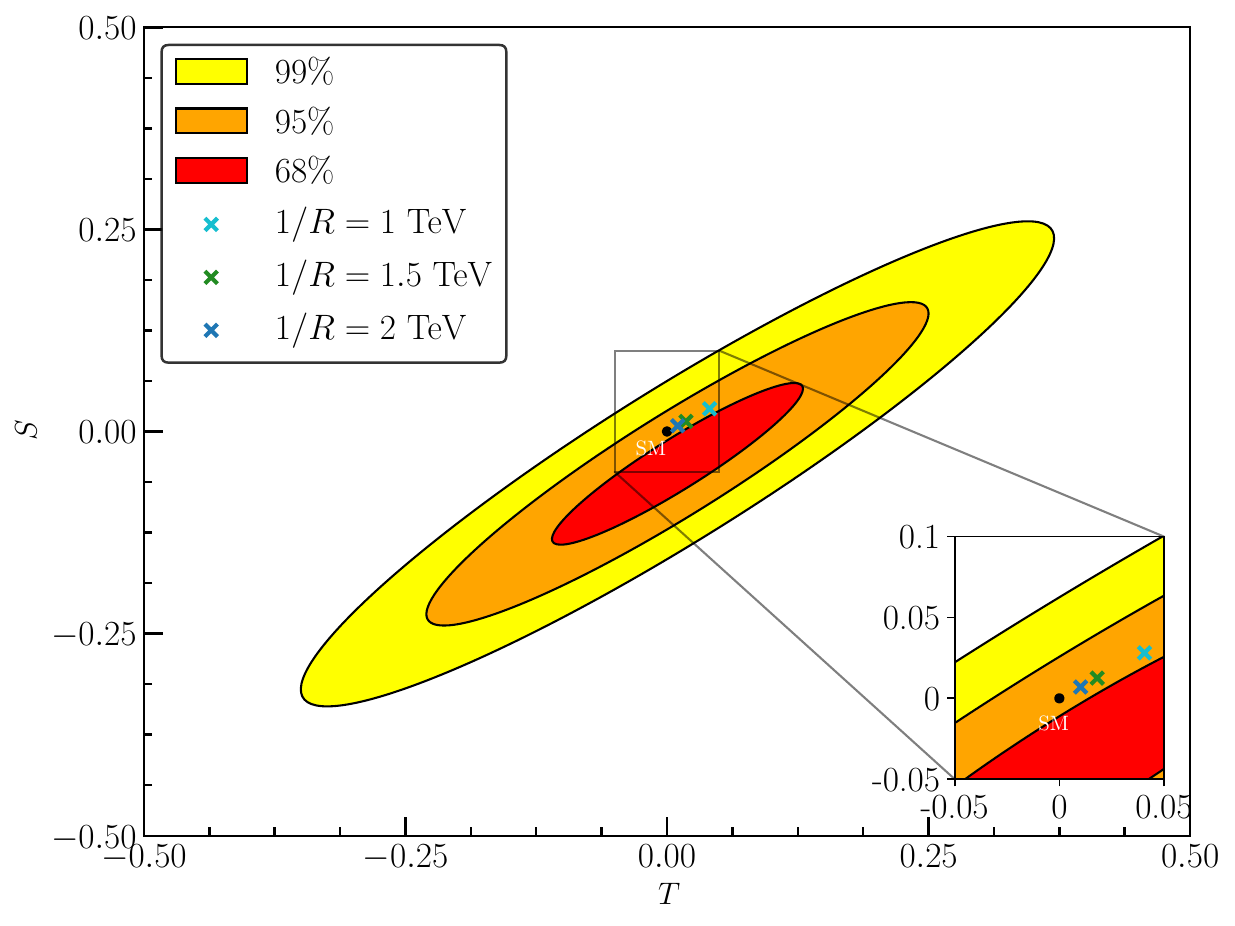}  
        \end{minipage}
        \begin{minipage}{0.5\textwidth}
            \centering
            \includegraphics[width=\textwidth]{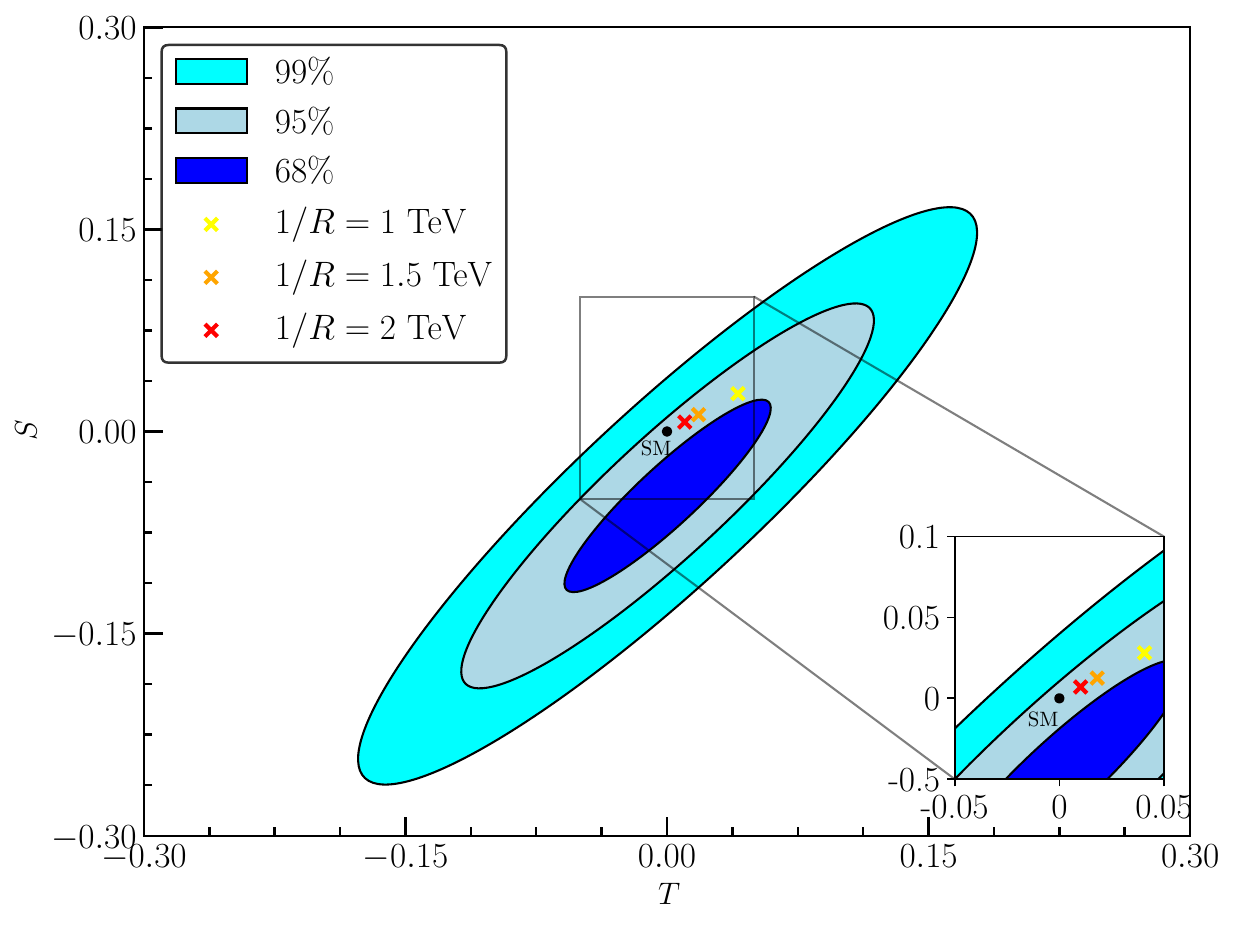}
        \end{minipage}
    \end{adjustbox}
    \caption{The confidence ellipses at $1\sigma$, $2\sigma$ and $3\sigma$ of the oblique parameters plotted in the $S,T$ plane for the experimental values in Eq.(\ref{STU_SM}). On the left $U$ is marginalised, while on the right it is set to $0$. We also show the aGUT predictions for three values of $1/R = 1,\ 1.5,\ 2$~TeV.}
    \label{SM_ST}
\end{figure}

As $U$ is typically sub-leading, it is customary to plot the results of precision electroweak measurements in a two-parameter space defined by $S$ and $T$, while $U$ is either marginalised or set to zero, the latter being more constraining for the theory. For current bounds, we can see in Figure \ref{SM_ST}  two different plots, one with $U$ free and one with $U=0$. Comparing with the aGUT predictions (indicated by crosses), we see that current precision does not constrain the model, as values of $1/R$ as low as $1$~TeV are still well within the 95\% region (two-$\sigma$). 
This is finally due to the fact that the corrections align to the flat direction of the current precision measurements.

\begin{figure}[ht!]
    \centering
    \begin{adjustbox}{center}
        \begin{minipage}[t]{0.5\textwidth}
            \centering
            \includegraphics[width=\textwidth]{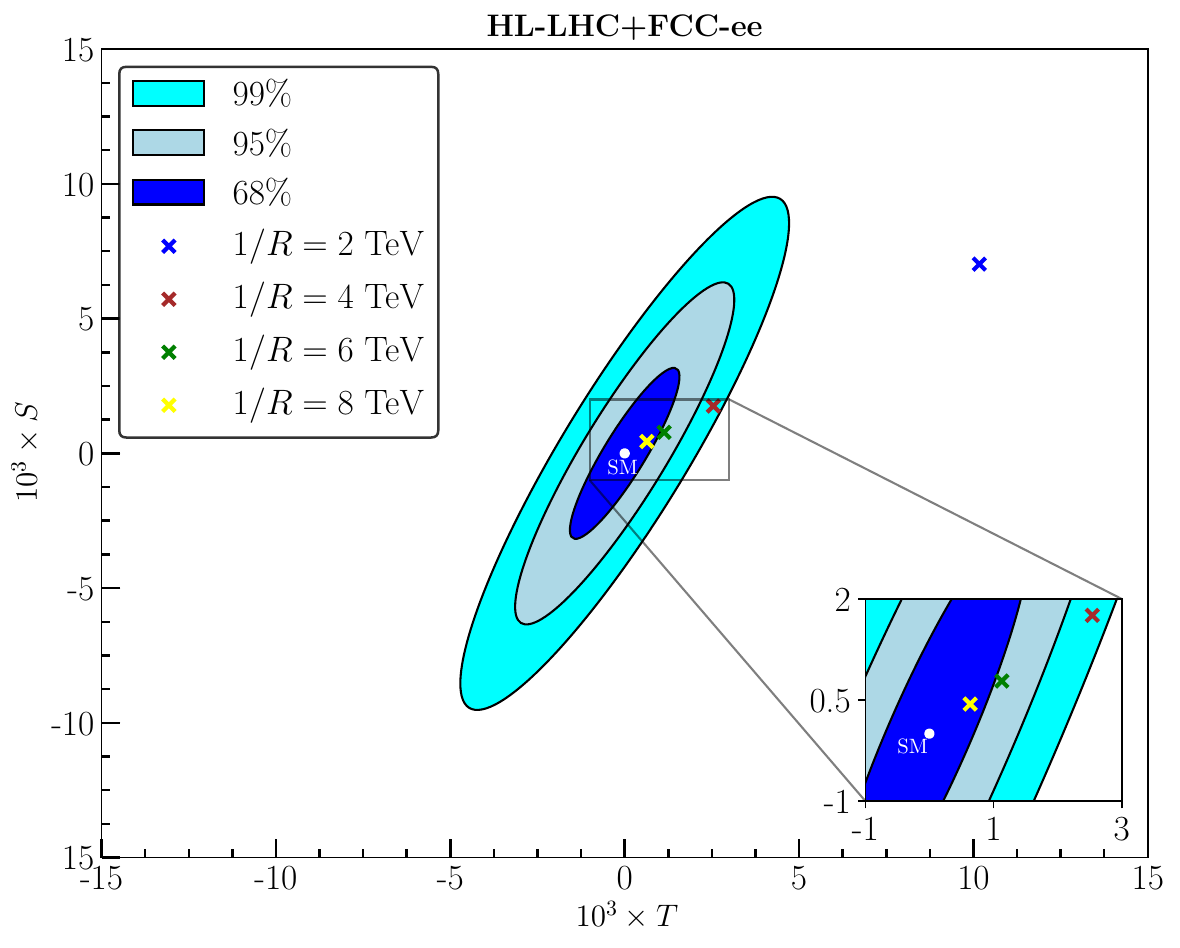}
        \end{minipage}
        \begin{minipage}[t]{0.5\textwidth}
            \centering
            \includegraphics[width=\textwidth]{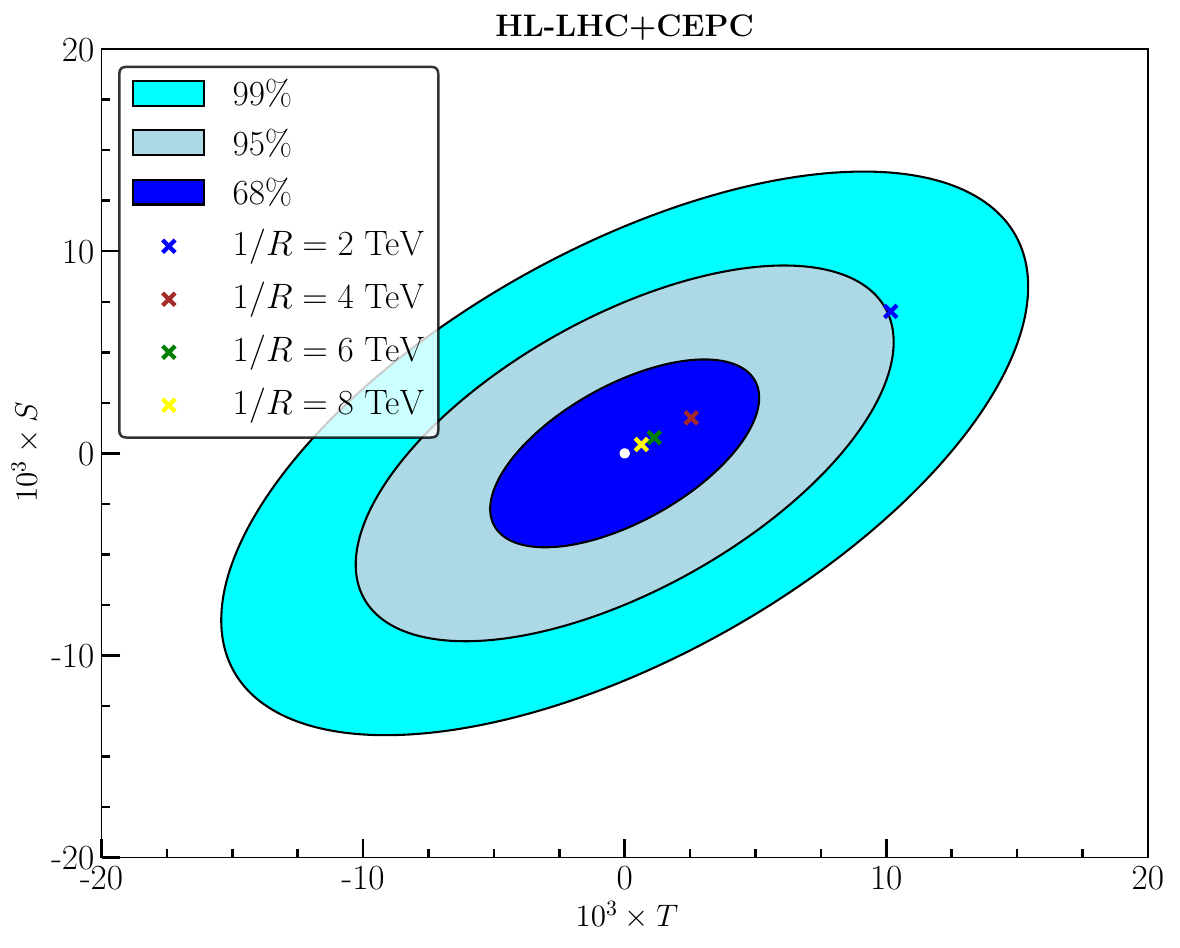}
        \end{minipage}
    \end{adjustbox}
    \caption{The confidence ellipses plotted in the $S,T$ plan at $1\sigma$, $2\sigma$ and $3\sigma$ for the prediction of the oblique parameters measurement Eq.(\ref{STU_FCC_CEPC}) of the FCC-ee (left) and CEPC (right) with $U=0$. The correlations are read off from Fig. 17 in \cite{deBlas:2019rxi}. Different values of $M= 1/R$ are considered to show the allowed values by the ellipse, the threshold of 95\% is reached by approximately 4 (2) TeV for FCC-ee (CEPC) respectively.}
    \label{FCC_ST}
\end{figure}

The projections for future colliders, using the values in  Eq.~\eqref{STU_FCC_CEPC}, are shown in Figure \ref{FCC_ST}, compared to various benchmarks of the $SU(5)$ aGUT model. We see that the level of precision reachable at FCC-ee and CEPC will be able to pose a non-trivial constraint on the model, also due to the different alignment of the flat direction. For FCC-ee, values of $1/R \lesssim 4$~TeV will be excluded at 95\% confidence level, while CEPC will be able to exclude values up to $\sim 2$ TeV. This bound is important as it can probe values of the compactification scales where the Indalo states can play the role of asymmetric Dark Matter, $1/R \sim 2.4$~TeV \cite{Cacciapaglia:2020qky}.

The KK states can also be directly produced at the LHC. The first tier, which contains $1$-modes with mass $M_1 = 1/R$, only contains Indalo states, which cannot decay into SM final state. Hence, they decay into the lightest Indalo, which generates a large missing energy. However, the mass splitting among these states is small, being induced at one loop level \cite{Cheng:2002iz}, hence leading to a hard-to-detect compressed spectrum (see, e.g., \cite{Edelhauser:2013lia}). A recast of Run-II data is likely to lead to an upper bound on $1/R$ between $1.5$ and $2$ TeV \cite{Deutschmann:2017bth,Avnish:2020atn,Flores:2021xwx}. On the other hand, the second tier, with masses $M_2 = 2/R$, can be singly produced and decay into SM states, both processes mediated at one loop \cite{Datta:2005zs}. The most promising signal stems from the lightest neutral vector state decaying into a pair of light leptons \cite{Cacciapaglia:2012dy,Edelhauser:2013lia}. The LHC reach for this channel could exceed $2$ TeV for the mass scale $1/R$. We explore this possibility in a next section in approximated way and a more detailed study would be needed to determine the reach of the LHC, which we leave for future studies. All considered, electroweak precision after FCC-ee and/or CEPC will have a better reach on $1/R$ than LHC. A future $100$~TeV hadron collider will have the ability to reach larger mass scales, at levels similar to the projected electroweak precision reach from Fig.~\ref{FCC_ST}.

\section{Collider bounds in the dilepton final states}
\begin{figure}[h!]
    \centering
    \includegraphics[width=0.35\linewidth]{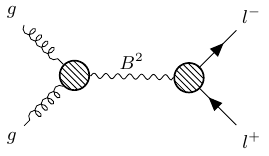}
    \includegraphics[width=0.35\linewidth]{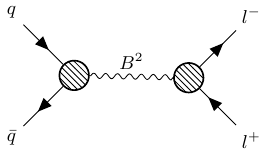}
    \caption{Direct production of the lightest neutral vector of the model $B^2_\mu$ decaying into two leptons. The explicit loops for the $qq$, $gg$ production and $B^2$ decay are given in Figures \ref{sect4:fig4} and \ref{sect4:fig5}. For simplicity, the model is limited to the 3rd generation of fermions, but similar interaction can be obtained for the two other generations.}
    \label{sect4:fig1}
\end{figure}

Depending on the compactification radius $R$, KK-modes could be seen in collider experiment with resonant and non resonant impact on cross sections. Sequential $Z'$ searches were made by CMS for a heavy neutral vector with SM coupling in the dilepton final states \cite{CMS:2021ctt} and their result can be recasted for extracting a bound on our model. In our case, since the initial states are SM particles i.e. 0-modes in terms of the KK-parity, only a 2-mode can be produced as a resonance from proton proton collision with an induced coupling of 1-mode particles to the 0-mode particles. The KK masses receive EW-conserving loop contributions from loops of KK modes. The form of the corrections is
\begin{equation}
\delta m_n = M_n \frac{Cg^2}{8\pi^2}\log\frac{\mu}{M_n}
\end{equation}
where $g$ is their typical coupling and $C$ a coefficient coming from all particle in the loop. As $g_S \gg g_2 \gg  g'$ the $U(1)_Y$ gauge field $B^n$ will get the smallest correction, hence $B^2$ is the lightest 2-mode. This correction typically dominates over the corrections coming from the Higgs vacuum expectation value, so we consider $B^n$ and $W^{3,n}$ as approximate mass eigenstates. Both the 2-modes and the 1-modes receive loop corrections, as such corrections are typically proportional to the mass, most decays of the form $2\rightarrow 1+1$ are kinematically forbidden. Only in some rare instances they may be open, and suppressed by the small phase space. So, typically, decays to two 1-modes are forbidden or very suppressed. For an estimate, we can neglect them and focus on the decays $2 \rightarrow 0+0$ through the induced couplings. The diagrams for the direct production and decay of $B^2$ are drawn in Figure \ref{sect4:fig1}. 
\\However, as $B^2$ has a small coupling to the colored particles of the loop, we expect the cross-section of this process to be negligible compared to the one producing a 2-mode of the gluon $g^2$. And since $B^2$ is the lightest 2-mode, it will always be present in the decay of $g^2$ and thus $\mathrm{BR}(g^2 \rightarrow X+ B^2)=1$.
\begin{figure}[h!]
    \centering
    \includegraphics[width=0.5\linewidth]{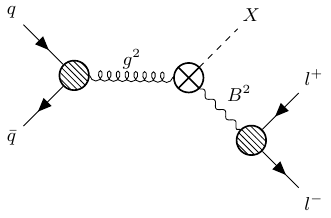}
    \caption{Main production channel for $B^2$ through the production of a 2-mode gluon $g^2$, $B^2$ being the lightest 2-mode we expect $\mathrm{BR}(g^2 \rightarrow X+ B^2)=1$. The two blobs are the same as for Figure \ref{sect4:fig1}, whereas the crossed dot represent the decay of $g^2$ into $B^2$, where a 0-mode X is produced.}
    \label{sect4:fig2}
\end{figure}
\\To estimate the cross-section $\sigma_{loop}(q\bar{q} \rightarrow g^2)$ we use the result of \cite{March:2005lxa} where they estimate $\sigma_{tree}(q\bar{q} \rightarrow g^*)$ of a heavy gluon (denoted $g^*$) assuming a SM coupling with quarks (with an additional $\sqrt{2}$ factor). Roughly we estimate the cross-section to be
\begin{equation}
    \sigma_{loop}(q\bar{q}\rightarrow g^2) = \frac{|F_{loop}|^2}{2g_S^2}\sigma_{tree}(q\bar{q} \rightarrow g^*),
\end{equation}
where $|F_{loop}|^2$ is a factor coming from the loops. The typical loops producing a 2-mode gluon are drawn Figure \ref{sect4:fig4} and contain a quark or a lepton and two leptoquarks. We estimate this loop factor as
\begin{equation}
    F_{loop} \sim C_{color}\frac{1}{16\pi^2}\prod_{i}g_i
\end{equation}
where $C_{color}$ is a color factor that we take equal to 3 and $g^i$ the three couplings present in the loop, in this case $\prod_i g_i = g_5^2g_S$ where the $SU(5)$ coupling $g_5$ appears at each vertex containing a leptoquark as gauge boson. Its canonical value obtained in standard GUT analyses is $\sim 0.7$, but without the precise running of the coupling we can't predict its precise value at the SM scale and take it as $0.5\lesssim g_5\lesssim0.8$, while the rest of the couplings are fixed at the SM scale $m_Z$. With this approximation we have
\begin{equation}
       \sigma_{loop}(q\bar{q}\rightarrow g^2) \simeq \frac{9}{2}\frac{g_5^4}{(16\pi^2)^2}\sigma_{tree}(q\bar{q} \rightarrow g^*).
\end{equation}
The cross-section for production of dileptons through the resonance $B^2$ is given by
\begin{equation}
\sigma(q\bar{q}\rightarrow X+B^2\rightarrow X + ee, \mu\mu) =\sigma_{loop}(q\bar{q}\rightarrow g^2)\mathrm{BR}(B^2 \rightarrow ee,\mu\mu).
\end{equation}
The branching ratio is simply
\begin{equation}
    \mathrm{BR}(B^2 \rightarrow ee,\mu\mu) = \frac{\Gamma_{ee,\mu\mu}}{\Gamma_{tot}} \simeq \frac{|{\cal{M}}_{ee,\mu\mu}|^2}{|{\cal{M}}_{tot}|^2},
\end{equation}
where each amplitude $\cal{M}$ originates from the sum of loop diagrams, as previously we evaluate each amplitude to be of the form
\begin{equation}
{\cal{M}}= \sum_{i\in loop} {\cal{M}}_i \simeq \frac{3}{16\pi^2}\sum_{i\in loop}\prod_{j\in vertex}g_j.
\end{equation}
For a simple approximation, we take into account only physical degrees of freedom and thus do not count loops with ghost, Goldstone bosons or gauge scalars. In addition, we include the two other fermion generations, which add a factor three in front of the associated amplitudes:
\begin{equation}
\begin{aligned}
16\pi^2{\cal{M}}_{\text{vectors}} =& g'^3\left[ \left(\frac{2}{3}\right)^3 + \left(-\frac{1}{3}\right)^3 + 2\left(\frac{1}{6}\right)^3  + 2\left(-\frac{5}{6}\right)^3\right] + 2 g'\left(\frac{1}{6} - \frac{5}{6}\right)\left(\frac{g_2}{2}\right)^2\\
&+2 g'\left(-\frac{5}{6}\right)\left(\frac{g_2}{\sqrt{2}}\right)^2+4\frac{g'}{6}\left(\frac{g_2}{\sqrt{2}}\right)^2\\
16\pi^2{\cal{M}}_{\text{quarks}} =& -3\times 5g'g_5^2
\\16\pi^2{\cal{M}}_{\text{leptons}} =& -3\times\frac{10}{3}g'g_5^2.
\end{aligned}
\end{equation}
\begin{figure}[hb!]
    \centering
    \includegraphics[width=0.75\linewidth]{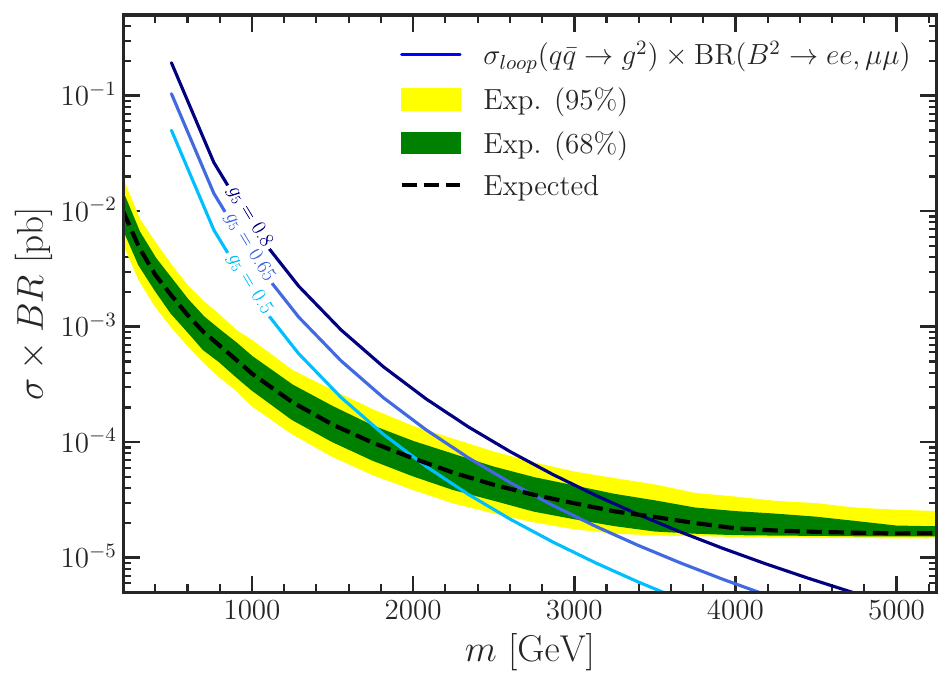}
    \caption{In yellow and green the upper limits at 95\% CL on the product of the production cross-section and the
branching fraction for a spin-1 resonance with a width equal to 0.6 of the resonance mass, relative to the product of the production cross-section and the branching fraction of a Z boson, multiplied
by the theoretical value of $\sigma(pp\rightarrow Z + X \rightarrow ll+ X)$ of 1928 pb. In blue the different theoretical values for the product $\sigma(q\bar{q}\rightarrow X+B^2\rightarrow X + ee, \mu\mu)$ depending on the value of $g_5$. The lightest line cross the expected line around 2 TeV, while the darkest one cross around 3 TeV, thus indicating that the experimental values for the cross-section excludes level-2 KK-mass values $2/R$ below 2 or 3 TeV, giving a bound on $1/R \sim 1.5$ TeV which is less strong than the EW constraints from Figure \ref{FCC_ST}.}
    \label{sect4:fig3}
\end{figure}
The comparison between the experimental bound from \cite{CMS:2021ctt}, can be seen in Figure \ref{sect4:fig3}. The details of the couplings and processes entering the calculation are given in Appendix \ref{sec:b2}. In Figure \ref{sect4:fig3} the different solid blue line crosses the expected cross-section at around 2 and 3 TeV, meaning that below this value the theoretical cross-section is large enough to be excluded by the experimental result. Above, the precision of the experimental value does not exclude the theoretical cross-section thus giving us a bound of $1/R \sim 1.5$ TeV for the compactification radius as the mass exclusion for the level-2 modes is about $\sim 3$ TeV. At first sight, exclusion from direct search look less significant than the electroweak precision calculations. However, a more detailed study is needed for a precise result, including the mass correction to the different 2-modes and the exact computation of the various amplitudes.

\section{Conclusions} \label{sec:concl}

We computed the \textit{oblique parameters} in the framework of a minimal $SU(5)$ aGUT model, based on one compactified extra dimension of radius $R$. We showed that in the limit where we the natural masses of the bottom quark and its Indalo partner are neglected, the two Kaluza-Klein towers merge together to give the expected result. In addition, we observed that the contribution of the new odd vectors and scalars give only a negative contribution to $S$ and no contribution to both $T$ and $U$. As expected, the $T$ parameter is the most constraining quantity, with contributions coming mainly from loops involving the Kaluza-Klein modes of the top quark. Finally, we made a comparison with the experimental values for different values of $R$, showing that the present results have limited sensitivity to the model. Projections for future collider experiments, such as FCC-ee and CEPC, instead, show sensitivity to Kaluza-Klein masses up to 4  and 2 TeV, respectively. These values bypass the expected reach at the LHC, although a detailed model analysis would be needed to be more quantitative. In addition to the EW precision test, we added the beginning of a study for the direct search of a 2-mode resonance in the dilepton final states in proton-proton collision at LHC using CMS results. At first glance, the resulting bound is less significant than the electroweak projection for FCC-ee, even though a more detailed study is needed.

We finally remark that these calculations can be used for other new physics models. The detailed computation of the involved Passarino-Veltman functions used are all given in appendix and can be reused easily, notably, for other models with one extra dimension, in particular other aGUT models based on larger gauge groups. Oblique parameters remain a very efficient and generic way of constraining models of new physics.

\newpage

\appendix
\section{Minimal SU(5) aGUT Model } \label{App A}
The gauge bosons $A^a_{M}$ and Grassmanian fields $\eta^a$ are embedded in the adjoint representation of $SU(5)$:
\begin{equation}
\mathbf{A}_{M} = \sum_{a=1}^{24}A^{a}_{M}T^{a} = {\begin{pmatrix}
\textbf{G}_{ij} & \textbf{X}_i/\sqrt{2} & \textbf{Y}_i/\sqrt{2}\\
\textbf{X}^\dagger_{i}/\sqrt{2}  & W^{3}/2 & W^{+}/\sqrt{2} \\
\textbf{Y}^\dagger_{i}/\sqrt{2}  & W^{-}/\sqrt{2} & -W^{3}/2 \\
\end{pmatrix}}_{M} + \sqrt{\frac{3}{5}}B_{M}\begin{pmatrix}
-\frac{1}{3}{\mathbbm{1}}_{3\times3} & 0 \\
0 & \frac{1}{2}{\mathbbm{1}}_{2\times2}
\end{pmatrix} 
\end{equation}
\begin{equation}
\boldsymbol{\eta} = \eta^aT^a =\begin{pmatrix}
\boldsymbol{\eta}_G-\dfrac{1}{3}\sqrt{\dfrac{3}{5}}\eta_B\mathbbm{1}_{3\times3} & \begin{pmatrix}
\boldsymbol{\eta}_X ^- & \boldsymbol{\eta}_Y^-
\end{pmatrix}/\sqrt{2}\\
\begin{pmatrix}
\boldsymbol{\eta}_X^+\\
\boldsymbol{\eta}_Y^+
\end{pmatrix}/\sqrt{2} & \boldsymbol{\eta}_W+\dfrac{1}{2}\sqrt{\dfrac{3}{5}}\eta_B\mathbbm{1}_{2\times2} 
\end{pmatrix}.
\label{AppA:eq-1}
\end{equation}
To define the gauge fixing part of the Lagrangian in a generic way, the matrix $5\times24$ $\mathbf{F}$ is defined as
\begin{equation}
\mathbf{F}= F^aT^a\qq{,} F^a = -iT^a\phi_v \qq{with} \phi_v = g\begin{pmatrix}
0 \\ 0 \\ 0 \\ 0 \\ \frac{v}{\sqrt{2}}
\end{pmatrix}\qq{and}g\phi_5 = \phi_v+\chi_5\,.
\label{AppA:eq0}
\end{equation}
As explained in \cite{PeskinSchroeder}, it is more convenient to use a real basis for the scalar fields of the Higgs multiplet; in this basis, written with a prime, $\mathbf{F}'$ is a sparse $10\times 24$ matrix and $\chi_5'$ a ten-component vector.
The scalars receive a mass contribution from different origins: the Higgs vacuum expectation value (vev), the KK masses, the gauge-fixing and the fifth component of the covariant derivative. Altogether, the different mass matrices have the form
\begin{equation}
\begin{aligned}
&\frac{1}{2}
\begin{pmatrix}
Z_5 & \phi_0 \\
\end{pmatrix}
\begin{pmatrix}
-\xi{M_n}^2-m_Z^2 & -\left(1-\xi\right)m_ZM_n\\
-\left(1-\xi\right)m_ZM_n & -{M_n}^2-\xi m_Z^2\\
\end{pmatrix}
\begin{pmatrix}
Z_5 \\ \phi_0
\end{pmatrix}
\\&+
\begin{pmatrix}
W_5^- & \phi^- \\
\end{pmatrix}
\begin{pmatrix}
-\xi{M_n}^2-m_W^2 & \left(1-\xi\right)m_WM_n\\
\left(1-\xi\right)m_WM_n & -{M_n}^2-\xi m_W^2\\
\end{pmatrix}
\begin{pmatrix}
 W_5^+ \\
 \phi^+
\end{pmatrix}\\
&+\begin{pmatrix}
\mathbf{Y}_5^\dagger & \mathbf{H}^\dagger\\
\end{pmatrix}
\begin{pmatrix}
-\xi M_n^2-m_W^2 & -\left(1-\xi\right)m_WM_n\\
-\left(1-\xi\right)m_WM_n & -{M_n}^2-\xi m_W^2\\
\end{pmatrix}
\begin{pmatrix}
\mathbf{Y}_5 \\
 \mathbf{H}
 \end{pmatrix}.
 \end{aligned}
 \label{AppA:eq4}
\end{equation}
 We notice that for $\xi = 1$ in the Feynman gauge, the matrices are diagonal. When working in a different gauge, where $\xi \neq 1$, these matrices can be diagonalised; defining $\tan\theta_{n}^{W,Z} = \frac{m_{W,Z}}{M_{n}}$ we get
\begin{equation}
\begin{aligned}
&
\begin{pmatrix}
{W^+_5}'\\
{\phi^+}'
\end{pmatrix} =\begin{pmatrix}
    c^W_n & s_n^W \\
    -s_n^W & c_n^W
\end{pmatrix}
\begin{pmatrix}
{W^+_5}\\
{\phi^+}
\end{pmatrix}
\qq{with}M' = \begin{pmatrix}
    -\xi \left(M^W_n\right)^2 & 0 \\
    0 & -\left(M^W_n\right)^2 
\end{pmatrix}\,,\\&
\begin{pmatrix}
{Z_5}'\\
{\phi^0}'
\end{pmatrix}=
\begin{pmatrix}
    c^Z_n & -s_n^Z \\
    s_n^Z & c_n^Z
\end{pmatrix}
\begin{pmatrix}
{Z_5}\\
{\phi^0}
\end{pmatrix}
\qq{with}M' = \begin{pmatrix}
    -\xi \left(M^Z_n\right)^2 & 0 \\
    0 & -\left(M^Z_n\right)^2 
\end{pmatrix}\,,\\&
\begin{pmatrix}
\mathbf{Y}_5'\\
\mathbf{H}'
\end{pmatrix}=\begin{pmatrix}
    c^W_n & -s_n^W \\
    s_n^W & c_n^W
\end{pmatrix}
\begin{pmatrix}
\mathbf{Y}_5\\
 \mathbf{H}
\end{pmatrix}\qq{with}M' = \begin{pmatrix}
    -\xi \left(M^W_n\right)^2 & 0 \\
    0 & -\left(M^W_n\right)^2 
\end{pmatrix}.
\end{aligned}
\label{AppA:eq5}
\end{equation}
Hence, in each sector one of the two scalars has a mass directly proportional to $\xi$, thus decoupling in the unitary gauge.

After rotation to the mass basis, the fermions denoted $f'$ and $q'_f$ are no longer sorted into singlets and doublets, and their interaction terms change. For odd fields, the charge conjugation can be removed using the properties of the charge conjugation operator
\begin{equation}
    \psi^c=C\overline{\psi}^T\qq{,}C^{-1}\gamma_{\mu}C= -\gamma_{\mu}^T\qq{and}C\gamma_5C^{-1}=\gamma_5^T.
    \label{AppB:eq3}
\end{equation}
All interactions between the 0-mode photon and fermions have the usual form
\begin{equation}
\bar{f}'eQ_f\slashed{A}f',\quad{\bar{q}_f'}eQ_f\slashed{A}q_f'.
 \label{AppB:eq4}
\end{equation}
For the $Z$ interactions, we denote by $I_f$ the isospin of the fermion $f$, then the mass mixing induces slightly more complicated vertices
\begin{equation}
\begin{aligned}
&\frac{g}{2c_w}\left(2I_f(s^f_n)^2-2Q_fs^2_w\right)\bar{f}'\slashed{Z}f'\;,\\
&\frac{g}{2c_w}\left(2I_f(c^f_n)^2-2Q_fs^2_w\right)\bar{q}'_f\slashed{Z}q'_f\;,\\
&\frac{g}{c_w}I_fs^f_nc^f_n\bar{f}'\slashed{Z}\gamma_5q'_f+\mathrm{h.c}\;.
\end{aligned}
 \label{AppB:eq5}
\end{equation}
\\The W interactions have no particular form since they mix doublet and singlet from different pairs, we illustrate here only the fields $(t,q_t),(b,q_b)$ and $(T,Q_t),(B,Q_b)$ that are involved in the loop computations:
\begin{equation}
\begin{aligned}
&\frac{g}{\sqrt{2}}\left(s^t_ns^b_n\overline{t}'\slashed{W}^+b'+c^t_nc^b_n\overline{q}_t'\slashed{W}^+q_b'-s^t_nc^b_n\overline{t}'\slashed{W}^+\gamma_5q_b'-s^b_nc^t_n\overline{q}_t'\slashed{W}^+\gamma_5b'\right)+\mathrm{h.c}\\
-&\frac{g}{\sqrt{2}}\left(s^t_ns^{\tau}_n\overline{T}'\gamma_{\mu}^T{W}^{\mu,-}{B}'+c^t_nc^{\tau}_n\overline{Q}_t'\gamma_{\mu}^T{W}^{\mu,-}Q_b'-s^n_tc^n_{\tau}\overline{T}'\gamma_{\mu}^T{W}^{\mu,-}\gamma_5^TQ_b'\right.\\&\left.\quad\quad-s^{\tau}_nc^t_n\overline{Q}'_t\gamma_{\mu}^T{W}^{\mu,-}\gamma_5^T{B}'\right)+\mathrm{h.c}.
\end{aligned}
 \label{AppB:eq6}
\end{equation}
\section{Passarino Veltman functions} \label{App B}
We define the Passarino-Veltman functions following the convention in \cite{Denner:1991kt,Denner:2019vbn}:
\begin{equation}
\begin{aligned}
&i\pi^2A_0(m) = \mu^{\epsilon}\int\dd[d]{k}\frac{1}{k^2-m^2},\\
&i\pi^2B_0(p^2,m_1,m_2)=\mu^{\epsilon}\int\dd[d]{k}\frac{1}{\left[k^2-m_1^2\right][(k-p)^2-m_2^2]},\\
&i\pi^2B_{\mu}(p^2,m_1,m_2) = \mu^{\epsilon}\int\dd[d]{k}\frac{k_{\mu}}{\left[k^2-m_1^2\right][(k-p)^2-m_2^2]} = i\pi^2p_{\mu}B_{1}(p^2,m_1,m_2),\\
&i\pi^2B_{\mu\nu} = \mu^{\epsilon}\int\dd[d]{k}\frac{k_{\mu}k_{\nu}}{\left[k^2-m_1^2\right][(k-p)^2-m_2^2]}\\&\quad\quad\quad\: =i\pi^2\left[B_{11}(p^2,m_1,m_2)p_{\mu}p_{\nu}+B_{00}(p^2,m_1,m_2)\eta_{\mu\nu}\right].
\end{aligned}
\end{equation}
For the computation of the oblique parameters, we are interested in two different cases. First, when $p^2=0$ for the $T$ parameter, and second, when the derivative of these functions with respect to $p^2$ is taken at $p^2 = 0$ for $S$ and $U$. Because we put the result in simpler form using the relations between the PV functions, we only need the specific formulas for $A_0$,$B_0$ and $B_{00}$. In the following, we give the definition of the PV function used in the text. We define
\begin{equation}
 \chi(x) = p^2x^2-(p^2+m_2^2-m^2_1)x+m_2^2,
\end{equation}
which is involved in the following definitions:
\begin{equation}
\begin{aligned}
&A_0(m) = m^2\left(1+\frac{1}{\bar{\epsilon}}-\ln\left(\frac{m^2}{\mu^2}\right)\right)\\
&B_0\left(p^2;m_1,m_2\right) = \frac{1}{\bar{\epsilon}} - \int_0^1 \dd{x}\ln\left(\frac{\chi}{\mu^2}\right),\\
&B_{00}(p^2,m_1,m_2) =\frac{1}{2}\left(\frac{1}{\bar{\epsilon}}+1\right)\int_0^1\dd{x}\chi - \frac{1}{2} \int_0^1 \dd{x}\chi \ln\left(\frac{\chi}{\mu^2}\right)\,.\\
\end{aligned}
\end{equation}
The convention for the UV part is
\begin{equation}
\frac{1}{\bar{\epsilon}} = \frac{2}{\epsilon}-\gamma+\ln4\pi\,.
\end{equation}
It is much simpler to apply the kinetic values, i.e $p^2 =0$ and/or $m_1 = m_2$, before performing the integration. Otherwise, indefinite spurious terms in $1/0$ can appear in the calculation. Here, except for the product $p^2B_0(p^2,m_1,m_2)$ that we evaluate at $p^2 = 0$ and where we need to have the analytic form of $B_0$, we do not need to perform the exact integration. The analytic form of $B_0$ reads as follows
\begin{equation}
\begin{aligned}
&B_0(p^2,m_1,m_2) = \frac{1}{\bar{\epsilon}}-\ln\left(\frac{m_2^2}{\mu^2}\right)-R+\frac{m_1^2-m_2^2+p^2}{2p^2}\ln\left(\frac{m_1^2}{m_2^2}\right)+2,\\
&\qq{with} R = -\frac{\Lambda}{p^2}\ln\left(\frac{-p^2+m_1^2+m_2^2+\Lambda}{2m_1m_2}\right)\qq{and}\Lambda^2 = \lambda(p^2,m_1^2,m_2^2)\,,
\end{aligned}
\end{equation}
where $\lambda$ is the Kallen function. The relevant combination $p^2B_0(p^2;m_1,m_2)$ is thus
\begin{equation}
\begin{aligned}
p^2B_0(p^2;m_1,m_2) =& p^2\left(\frac{1}{\bar{\epsilon}}-\ln\left(\frac{m_2^2}{\mu^2}\right)\right)+\Lambda\ln\left(\frac{-p^2+m_1^2+m_2^2+\Lambda}{2m_1m_2}\right)\\
&+\frac{1}{2}\left(m_1^2-m_2^2+p^2\right)\ln\left(\frac{m_1^2}{m_2^2}\right)+2p^2.
\end{aligned}
\end{equation}
In addition, we define two combinations of PV functions from \cite{BardinBible} that appear in the loop results:
\begin{equation}
\begin{aligned}
A^1_3(m) &= 3A_0(m)-2m^2\;,\\
A_1^1(p^2,m_1^2,m_2^2) &= 10B_{00}(p^2,m_1,m_2)+(m_1^2+m_2^2+4p^2)B_{0}\\&\quad\quad-A_0(m_1)-A_0(m_2)+2\left(m_1^2+m_2^2-\frac{p^2}{3}\right)\;.
\end{aligned}
\label{AppC:eq1}
\end{equation}
\subsection{Values at $p^2=0$ for the $T$ parameter}
When $p^2=0$ the $\chi$ variable becomes
\begin{equation}
    \chi = -(m_2^2-m_1^2)x+m_2^2 = m_1^2 x+(1-x)m_2^2\;,
\end{equation}
and when $m_1 = m_2$, we get the very simple but useful formulas
\begin{equation}
\begin{aligned}
&B_0(0,m,m) = \frac{1}{\bar{\epsilon}}-\ln\frac{m^2}{\mu^2}\;\,\\
&B_{00}(0,m,m) = \frac{1}{2}m^2\left(1+\frac{1}{\bar{\epsilon}}-\ln\frac{m^2}{\mu^2}\right)\;.
\end{aligned}
\end{equation}
When $m_1 \ne m_2$, $B_0$ reads
\begin{equation}
B_0(0,m_1,m_2) = \frac{1}{\bar{\epsilon}}-\int\dd{x}\ln\left(\frac{m_1^2x+(1-x)m_2^2}{\mu^2}\right),
\end{equation}
using the fact that $\int_0^1\dd{x}\ln\left(xa+(1-x)b\right)= \dfrac{a\ln a-b\ln b}{a-b}-1$, we arrive at
\begin{equation}
B_0(0,m_1,m_2) = \frac{A_0(m_1)-A_0(m_2)}{m_1^2-m_2^2}\,,
\end{equation}
or in an equivalent way
\begin{equation}
B_0(0,m_1,m_2) = \frac{1}{\bar{\epsilon}}-\ln\left(\frac{m_1m_2}{\mu^2}\right)+1-\frac{m_1^2+m_2^2}{2(m_1^2-m_2^2)}\ln\left(\frac{m_1^2}{m_2^2}\right)
\end{equation}
For $B_{00}$ the expression is
\begin{equation}
\begin{aligned}
B_{00}(0,m_1,m_2) &= \frac{1}{4}\left(m_1^2+m_2^2\right)\left(1+\frac{1}{\bar{\epsilon}}-\ln\frac{m_2^2}{\mu^2}\right)\\&-\frac{1}{2}\int_0^1\dd{x}\left(m_2^2-(m^2_2-m_1^2)x\right)\ln\left(\frac{m_2^2-(m_2^2-m_1^2)x}{m_2^2}\right),
\end{aligned}
\end{equation}
where we set $u = 1-\left(1-\dfrac{m_1^2}{m_2^2}\right)x$ and hence:
\begin{equation}
\begin{aligned}
B_{00}(0,m_1,m_2) &= \frac{1}{4}\left(m_1^2+m_2^2\right)\left(1+\frac{1}{\bar{\epsilon}}-\ln\frac{m_2^2}{\mu^2}\right)-\frac{m_2^4}{2(m_1^2-m_2^2)}\left[\frac{u^2}{4}\left(2\ln u-1\right)\right]_{u=1}^{u=\frac{m_1^2}{m_2^2}}\\
&=\frac{1}{4}\left(m_1^2+m_2^2\right)\left(\frac{3}{2}+\frac{1}{\bar{\epsilon}}-\ln\frac{m_2^2}{\mu^2}\right)-\frac{m_1^4}{4(m_1^2-m_2^2)}\ln\frac{m_1^2}{m_2^2}\:.
\end{aligned}
\end{equation}
It can also be useful to have a symmetric formula with respect to the masses:
\begin{equation}
\begin{aligned}
B_{00}(0,m_1,m_2) &=\frac{1}{4}\left(m_1^2+m_2^2\right)\left(\frac{3}{2}+\frac{1}{\bar{\epsilon}}-\ln\frac{m_1m_2}{\mu^2}\right)-\frac{m_1^4 + m_2^4}{8(m_1^2-m_2^2)}\ln\frac{m_1^2}{m_2^2}\:.
\end{aligned}
\end{equation}
For the product of $B_0$ with $p^2 = 0$, we have $\Lambda = \sqrt{(m_1^2-m_2^2)^2} = |m_1^2-m_2^2|$, hence
\begin{equation}
\begin{aligned}
\lim_{p^2\rightarrow 0}p^2B_0(p^2;m_1,m_2)=&|m_1^2-m_2^2|\ln\left(\frac{m_1^2+m_2^2-|m_1^2-m_2^2|}{2m_1m_2}\right)\\&+\frac{m_1^2-m_2^2}{2}\ln\left(\frac{m_1^2}{m_2^2}\right) = 0\;.
\end{aligned}
\end{equation}
In the case of Kaluza-Klein modes, the different masses can be written as $(M_n^i)^2 = (M_n)^2(1+z_i)$, where $z_i = m_i^2/M_n^2$ and $i=1,2$. For each previous functions, it is better to have an exact expression in terms of $z_i$, $\Delta = 1/\bar{\epsilon}+\ln (M_n)^2/\mu^2$ and the associated Taylor expansions:
\begin{equation}
\begin{aligned}
A_0(M_n^1) &= (M_n)^2(1+z_1)\left[1+\Delta-\ln(1+z_1)\right]\\
&= (M_n)^2(1+z_1)\Delta+(M_n)^2\left[1-\frac{z_1^2}{2}+\frac{z_1^3}{6}+{\cal{O}}\left(\frac{1}{(M_n)^6}\right)\right]\;,\\
B_{00}(0,M_n^1,M_n^1) &=\frac{1}{2}(M_n)^2(1+z_1)\left[\Delta+1-\ln(1+z_1)\right]\\
& = \frac{1}{2}(M_n)^2(1+z_1)\Delta + (M_n)^2\left[\frac{1}{2}-\frac{z_1^2}{4}+\frac{z_1^3}{12}+{\cal{O}}\left(\frac{1}{(M_n)^6}\right)\right]\;,\\
\end{aligned}
\end{equation}
\begin{equation}
\begin{aligned}
B_0(0,M_n^1,M_n^2) &= \Delta+1-\frac{1}{z_1-z_2}\left[(1+z_1)\ln(1+z_1)-(1+z_2)\ln(1+z_2)\right]\\
&=\Delta - \frac{z_2+z_1}{2}+\frac{z_1^2+z_1z_2+z_2^2}{6}+{\cal{O}}\left(\frac{1}{(M_n)^4}\right)\;,\\
B_{00}(0,M_n^1,M_n^2) &= (M_n)^2\left[\frac{1}{4}(2+z_1+z_2)\left(\frac{3}{2}+\Delta-\ln(1+z_2)\right)+\frac{(1+z_2)^2}{4(z_2-z_1)}\ln\left(\frac{1+z_1}{1+z_2}\right)\right]\\
&=(M_n)^2\left[\frac{2+z_1+z_2}{4}\Delta+\frac{1}{2}-\frac{z_1^2+z_1z_2+z_2^2}{12}\right.\\&\left.\quad\quad\quad\quad\quad + \frac{z_1^3+z_1^2z_2+z_1z_2^2+z_2^3}{48}+{\cal{O}}\left(\frac{1}{(M_n)^6}\right)\right]\;,
\end{aligned}
\end{equation}
\subsection{Derivatives at $p^2=0$ for the $S$ and $U$ parameters}
To compute $S$ and $U$ we need the derivatives of the Passarino-Veltman functions:
\begin{equation}
\begin{aligned}
    &\pdv{B_0}{p^2}\left(p^2,m_1,m_2\right) = -\int_0^1\dd{x}\frac{x(x-1)}{p^2x(x-1)+(m_1^2-m_2^2)x+m_2^2}\;,\\
    &\pdv{B_{00}}{p^2}\left(p^2,m_1,m_2\right) = -\frac{1}{12}\left(\frac{1}{\bar{\epsilon}}-\ln\frac{m_2^2}{\mu^2}\right)-\frac{1}{2}\int_0^1x(x-1)\ln\left(\frac{\chi}{m_2^2}\right)\:,
\end{aligned}
\end{equation}
in the case of a null momentum:
\begin{equation}
\begin{aligned}
&\pdv{B_0}{p^2}\left(0,m_1,m_2\right) = -\int_0^1\dd{x}\frac{x(x-1)}{m_2^2(1-x)+m_1^2x}\\&\quad\quad\quad\quad\quad\quad\quad=\frac{1}{2(m_1^2-m_2^2)^3}\left[m_1^4-m_2^4-2m_1^2m_2^2\ln\left(\frac{m_1^2}{m_2^2}\right)\right],\\
&\pdv{B_{00}}{p^2}\left(0,m_1,m_2\right) =-\frac{1}{12}\left(\frac{1}{\bar{\epsilon}}-\ln\frac{m_2^2}{\mu^2}\right)-\frac{1}{2}\int_0^1\dd{x}x(x-1)\ln\left(1+\frac{m_1^2-m_2^2}{m_2^2}x\right)\\
&\quad\quad\quad\quad\quad\quad =-\frac{1}{12}\left(\frac{1}{\bar{\epsilon}}-\ln\frac{m_2^2}{\mu^2}\right)\\&\quad\quad\quad\quad\quad\quad\quad+\frac{1}{72}\left[\frac{-5m_1^4+22m_1^2m_2^2-5m_2^2}{(m_1^2-m_2^2)^2}+\frac{6(m_1^6-3m_1^4m_2^2)}{(m_1^2-m_2^2)^3}\ln\left(\frac{m_1^2}{m_2^2}\right)\right].\\
\end{aligned}
\end{equation}
Again, $B_{00}$ can be put under a symmetric form as
\begin{equation}
\begin{aligned}
\pdv{B_{00}}{p^2}\left(0,m_1,m_2\right) &= -\frac{1}{12}\left(\frac{1}{\bar{\epsilon}}-\ln\frac{m_1m_2}{\mu^2}\right)-\frac{1}{72}\biggl[\frac{5m_1^4-22m_1^2m_2^2+5m_2^4}{(m_1^2-m_2^2)^2}\\&\quad\quad\quad-\frac{3(m_1^6-3m_1^4m_2^2-3m_1^2m_2^4+m_2^6)}{(m_1^2-m_2^2)^3}\ln\left(\frac{m_1^2}{m_2^2}\right)\biggr].\\
\end{aligned}
\end{equation}
In particular, for equal masses we have
\begin{equation}
\pdv{B_0}{p^2}\left(0,m,m\right)=\frac{1}{6m^2}.
\end{equation}
We also need the derivative of the product of $p^2$ and $B_0$ which reads
\begin{equation}
\begin{aligned}
\pdv{}{p^2}\left(p^2B_0(p^2,m_1,m_2)\right) &=\left(\frac{1}{\bar{\epsilon}}-\ln\frac{m_2m_1}{\mu^2}\right)+\frac{p^2-m_1^2-m_2^2}{\Lambda}\ln\left(\frac{-p^2+m_1^2+m_2^2+\Lambda}{2m_1m_2}\right)\\&+\frac{p^2-m_1^2-m_2^2-\Lambda}{p^2+m_1^2+m_2^2+\Lambda}+2,
\end{aligned}
\end{equation}
and which is simplified in the case of $p^2=0$
\begin{equation}
\begin{aligned}
 \pdv{}{p^2}\left(p^2B_0(p^2,m_1,m_2)\right)\eval_{p^2=0} =&  \left(\frac{1}{\bar{\epsilon}}-\ln\frac{m_2m_1}{\mu^2}\right)+1\\&-\frac{m_1^2+m_2^2}{|m_1^2-m_2^2|}\ln\left(\frac{m_1^2+m_2^2+|m_1^2-m_2^2|}{2m_1m_2}\right)\\=&\left(\frac{1}{\bar{\epsilon}}-\ln\frac{m_2^2}{\mu^2}\right)+1-\frac{m_1^2}{m_1^2-m_2^2}\ln\frac{m_1^2}{m_2^2}
 \end{aligned}
\end{equation}
where we can see that is symmetric under the exchange of $m_1 \leftrightarrow m_2$. In addition, when both masses are equal we find
\begin{equation}
\begin{aligned}
 \pdv{}{p^2}\left(p^2B_0(p^2,m,m)\right)\eval_{p^2=0} =& \frac{1}{\bar{\epsilon}}-\ln\frac{m^2}{\mu^2}+1+\lim_{p^2\rightarrow 0}\frac{p^2-2m^2}{\Lambda}\ln\left(1-\frac{p^2}{2m^2}+\frac{\Lambda}{2m^2}\right)\\=&\frac{1}{\bar{\epsilon}}-\ln\frac{m^2}{\mu^2}\;.
 \end{aligned}
\end{equation}
In our specific setup, with KK masses we have in terms of $z_i$
\begin{equation}
\begin{aligned}
 \pdv{}{p^2}\left(p^2B_0(p^2,m_1,m_2)\right)\eval_{p^2=0} &=\Delta +1-\ln(1+z_2)-\frac{1+z_1}{z_1-z_2}\ln\left(\frac{1+z_1}{1+z_2}\right)\\&=\Delta-\frac{z_1+z_2}{2}+\frac{z_1^2+z_1z_2+z_2^2}{6}+{\cal{O}}\left(\frac{1}{(M_n)^4}\right)\;,
\end{aligned}
\end{equation}
and
\begin{equation}
\begin{aligned}
\pdv{B_0}{p^2}\left(0,M_n^1,M_n^2\right) &= \frac{1}{2(M_n)^2(z_1-z_2)^3}\left[(1+z_1)^2-(1+z_2)^2-2(1+z_1)(1+z_2)\ln\left(\frac{1+z_1}{1+z_2}\right)\right]\\
&= \frac{1}{(M_n)^2}\left[\frac{1}{6}-\frac{z_1+z_2}{12}+\frac{z_1^2+z_2^2}{20}+\frac{z_1z_2}{15}+{\cal{O}}\left(\frac{1}{(M_n)^4}\right)\right]\;,\\
\pdv{B_{00}}{p^2}\left(0,M_n^1,M_n^2\right) &=-\frac{1}{12}\Delta+\frac{1}{12}\ln(1+z_2)-\frac{1}{12}\left[\frac{3(1+z_1)(1+z_2)^2-(1+z_2)^3}{(z_1-z_2)^3}-1\right]\ln\left(\frac{1+z_1}{1+z_2}\right)\\&\quad -\frac{5(1+z_1)^2-22(1+z_1)(1+z_2)+5(1+z_2)^2}{72(z_1-z_2)^2}\\
&= -\frac{1}{12}\Delta +\frac{z_1+z_2}{24}-\frac{z_2^2+z_1^2}{80}-\frac{z_2z_1}{60}+{\cal{O}}\left(\frac{1}{(M_n)^4}\right)\;.
\end{aligned}
\end{equation}
\newpage
\section{Fermionic loops} \label{App C}
\begin{figure}[ht!]
    \centering
    \vspace{1cm}
    \includegraphics[scale=1]{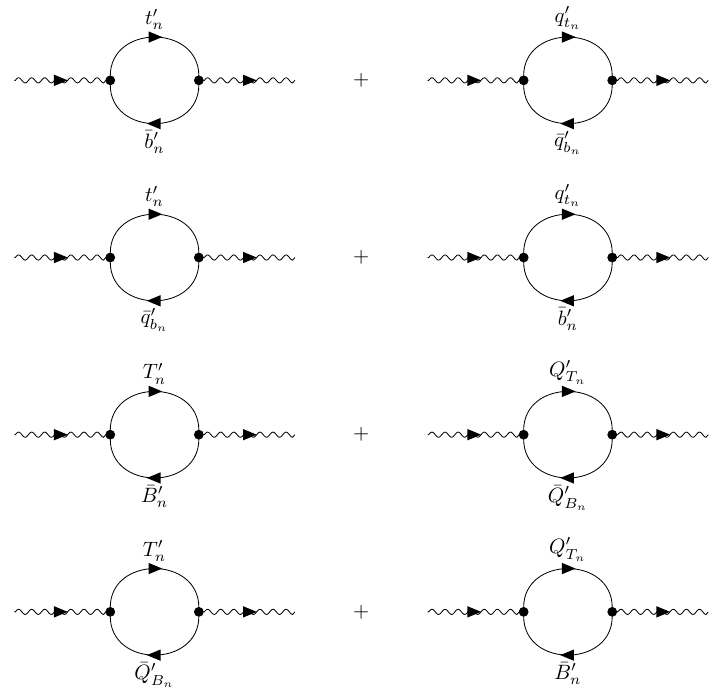}
    \caption{Loops involving the top quark mass in the $W^+,W^-$ self-energies, for loops with Indalo particles, $n$ is an odd number while for loops with KK $n$ is non-zero even number.}
    \label{fermionic_loops_WW}
\end{figure}

\begin{figure}[ht!]
    \centering
    \vspace{1cm}
    \includegraphics[scale=1]{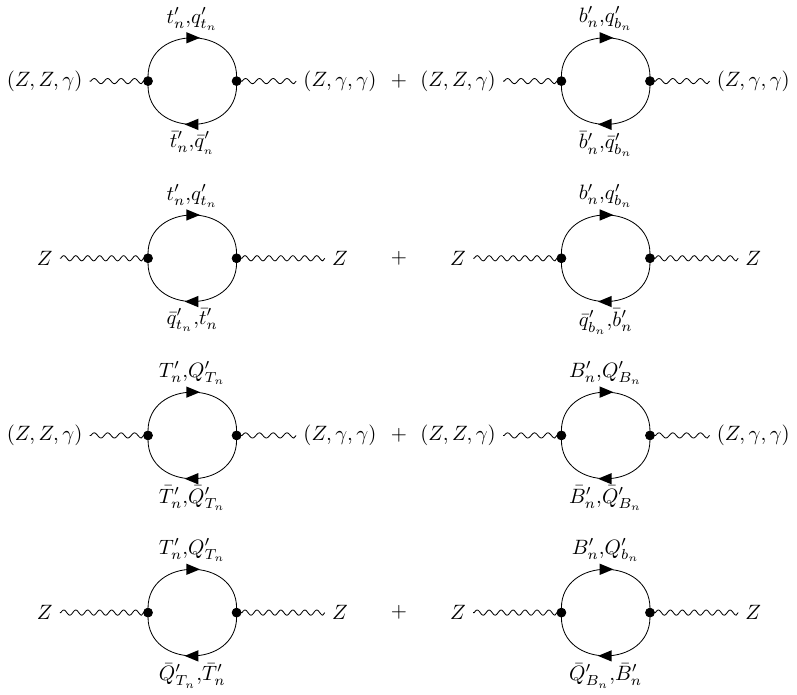}
    \caption{Loops involving the top quark mass in the $ZZ,Z\gamma,\gamma\gamma$ self energies, $n$ is a non-zero even number for KK particles and an odd number for Indalo particles.}
    \label{fermionic_loops_33}
\end{figure}

The self energy can then be written as
\begin{equation}
\begin{aligned}
&i\Pi_{\mu\nu} = -\int\frac{\mathrm{d}^4k}{\left(2\pi\right)^4}\mathrm{Tr}\left[i\gamma_{\mu}\left(v-a\gamma_5\right)\dfrac{i\left(\slashed{k}+m_1\right)}{k^2-m_1^2}i\gamma_{\nu}\left(v'-a'\gamma_5\right)\dfrac{i\left(\slashed{k}-\slashed{p}+m_2\right)}{\left(k-p\right)^2-m_2^2}\right].
\end{aligned}
\end{equation}
After doing traces over Dirac matrices, and defining $k^2-m_1^2 = d_1$, $\left(k-p\right)^2-m_2^2 = d_2$  we arrive at
\begin{equation}
\begin{aligned}
i\Pi_{\mu\nu}(p^2) =&- \int\frac{\mathrm{d}^4k}{{\left(2\pi\right)^4}}\frac{1}{d_1d_2}\biggl\{\left(aa'+vv'\right)(k-p)_{\rho}k_{\sigma}\mathrm{Tr}\left(\gamma_{\mu}\gamma_{\rho}\gamma_{\nu}\gamma_{\sigma}\right)+(vv'-aa')m_1m_2\mathrm{Tr}\left(\gamma_{\mu}\gamma_{\nu}\right)\biggr\}\\
=& -\frac{4i}{16\pi^2}\int\frac{\mathrm{d}^4k}{i\pi^2}\frac{1}{d_1d_2}\bigg\{(aa'+vv')(2k_{\mu}k_{\nu}-p_{\mu}k_{\nu}-p_{\nu}k_{\mu}-\eta_{\mu\nu}k^2+\eta_{\mu\nu} k\cdot p)\\&\quad\quad\quad\quad\quad\quad\quad\quad\quad\quad + (vv'-aa')m_1m_2\eta_{\mu\nu}\biggr\}.
\end{aligned}
\end{equation}
The $\eta_{\mu\nu} $ part can then be rewritten as
\begin{equation}
\begin{aligned}
i\Pi_{VV'}=&-\frac{4i}{16\pi^2}\biggl\{(aa'+vv')\left( -(d-2)B_{00} -p^2B_{11} +p^2B_1\right)\\&\quad\quad\quad\quad+ (aa'-vv')m_1m_2B_0\biggr\}\left(p^2,m_1,m_2\right)\\
&=\frac{2}{16\pi^2}\biggl\{(aa'+vv')\left(2dB_{00}+2p^2B_{11}-4B_{00}-2p^2B_1\right)\\&\quad\quad\quad\quad+2(aa'-vv')m_1m_2B_0\biggr\}\left(p^2,m_1,m_2\right).
\end{aligned}
\end{equation}
It is better to put these expressions under a form involving the simplest combination of PV functions. One can show that
\begin{equation}
\begin{aligned}
&p^2B_{11}(p^2;m_1,m_2)+ dB_{00}(p^2,m_1,m_2)= A_0(m_2)+m_1^2B_0(p^2;m_1,m_2),\\
&p^2B_1(p^2;m_1,m_2) = \frac{1}{2}\left[A_0(m_2)-A_0(m_1)+(p^2+m_1^2-m_2^2)B_0(p^2,m_1,m_2)\right],
\end{aligned}
\end{equation}
which enables us to write
\begin{equation}
\begin{aligned}
\Pi_{VV'} &= \frac{2}{16\pi^2}\biggl\{(aa'+vv')\left(A_0(m_2)+A_0(m_1)+(m_1^2+m_2^2-p^2)B_0-4B_{00}\right)\\&\quad\quad\quad\quad+\,2(aa'-vv')m_1m_2B_0\biggr\}
.\\
\end{aligned}
\end{equation}
Now we list the fermionic loops involved in the charged and neutral self-energies, the loops involved in the $W^+,W^-$ self-energies can be seen Fig. \ref{fermionic_loops_WW}, and the one for $ZZ,Z\gamma,\gamma\gamma$ Fig. \ref{fermionic_loops_33}.
\clearpage
\newpage
\section{Vector and scalar loops} \label{App D}
\begin{figure}[ht!]
\hspace*{-1.75cm}
    \centering
    \includegraphics[scale=0.9]{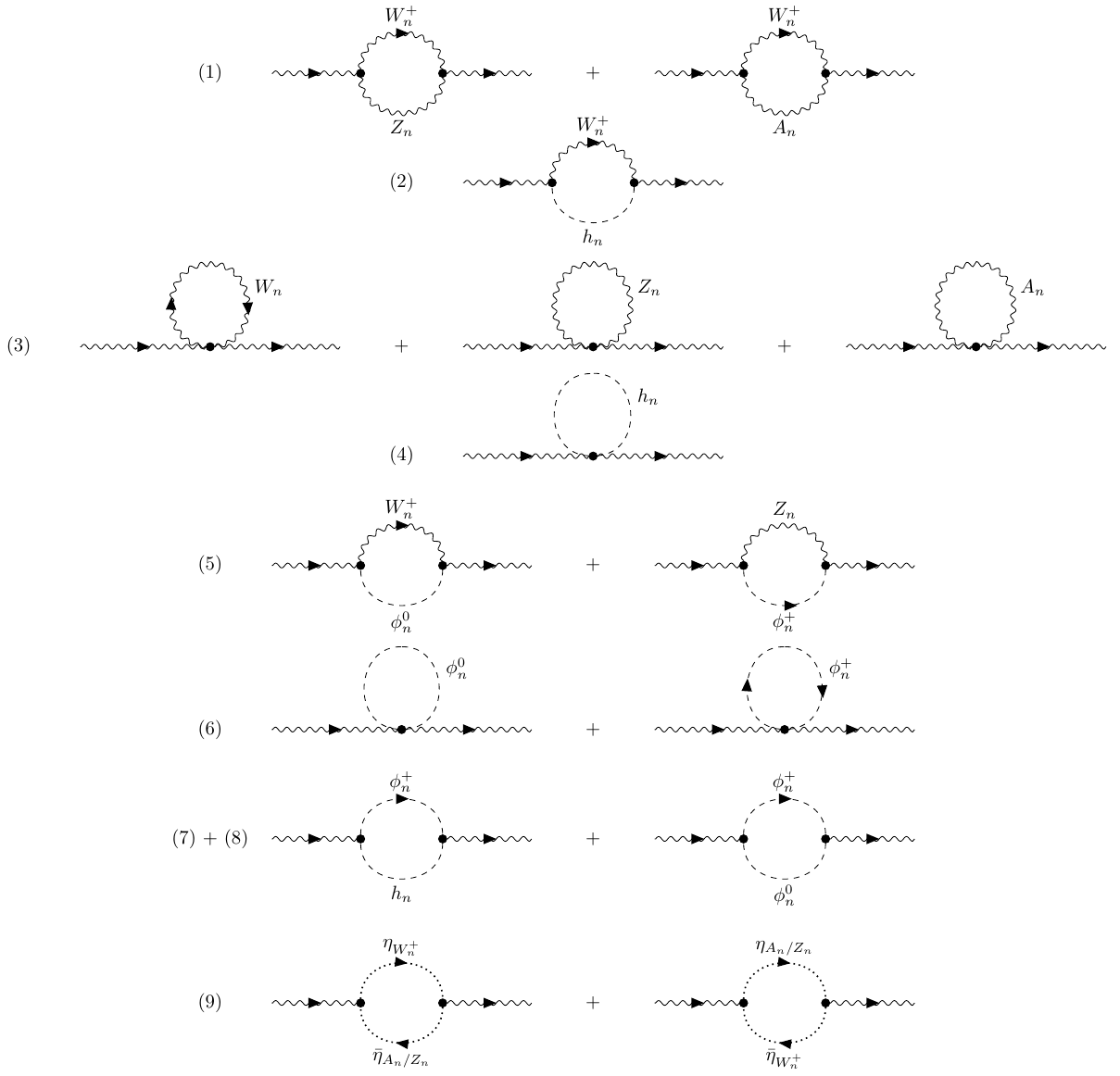}
    \caption{Loops involved in the $W$ self-energy for the SM-like subset of loops. The sum on all even Kaluza-Klein modes denoted $n$ has be to be taken afterward. The associated two point functions can be found Eq.(\ref{AppC:eq4}) }
    \label{fig5}
\end{figure}

\begin{figure}[ht!]
    \centering
    \includegraphics[scale=0.9]{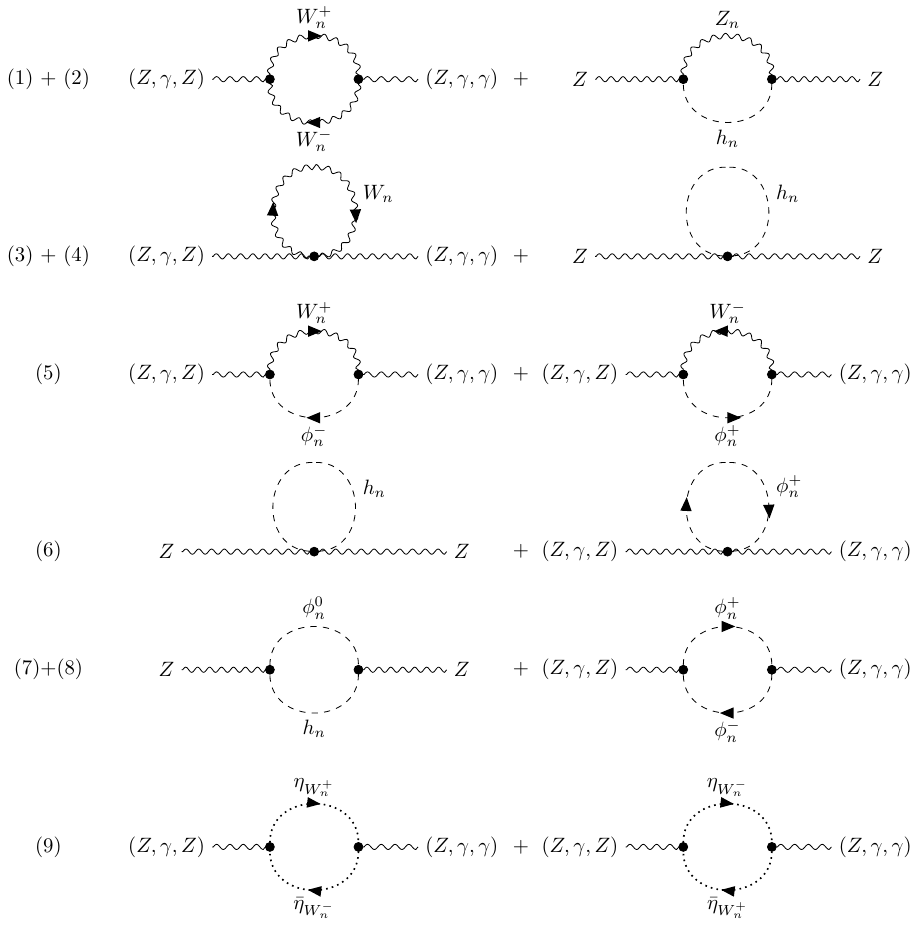}
    \caption{Loops involved in the $ZZ,\gamma\gamma,Z\gamma$ self-energy for the SM-like subset of loops. The sum on all even Kaluza-Klein modes denoted $n$ has be to be taken afterward. The associated two point functions can be found Eq.(\ref{AppC:eq5}) }
    \label{fig6}
\end{figure}

The scalars and vectors loops are divided in three sets, first the SM-like loops Figure \ref{fig5} and  Figure \ref{fig6} for the charged and neutral self-energies respectively.

\newpage

\begin{figure}[ht!]
    \centering
    \includegraphics[scale=0.9]{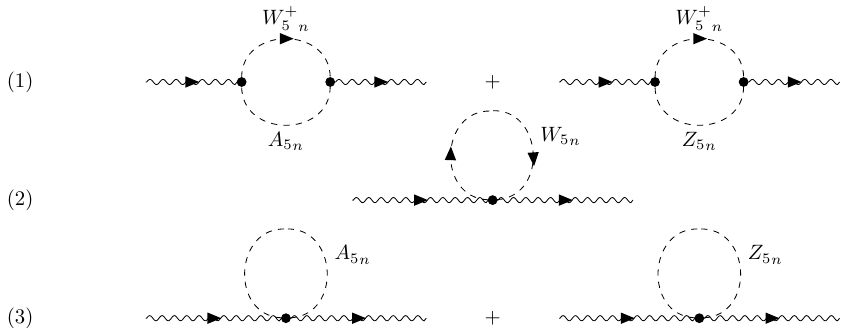}
    \caption{Loops involved in the $W$ self-energy for the new even subset of loops. The sum on all even Kaluza-Klein modes denoted $n$ has to be taken afterward. The associated two point functions can be found Eq.(\ref{AppC:eq6}) }
    \label{fig7}
\end{figure}

\begin{figure}[ht!]
    \centering
    \includegraphics[scale=0.9]{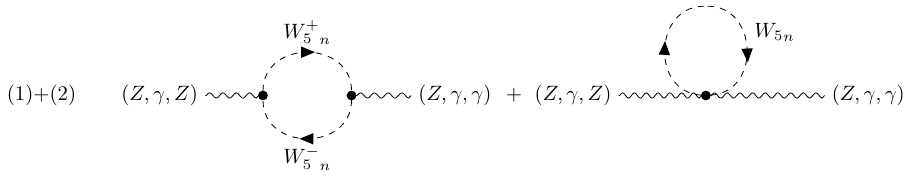}
    \caption{Loops involved in the  $ZZ,\gamma\gamma,Z\gamma$ self-energies for the new even subset of loops. The sum on all even Kaluza-Klein modes denoted $n$ has be to be taken afterward. The associated two point functions can be found Eq.(\ref{AppC:eq7}) }
    \label{fig8}
\end{figure}

\newpage
Then we have the new even loops, less numerous, in Figure \ref{fig7} and \ref{fig8}, which involve the gauge scalars present in non-unitary gauge.

\begin{figure}[ht!]
    \centering
    \includegraphics[scale=0.9]{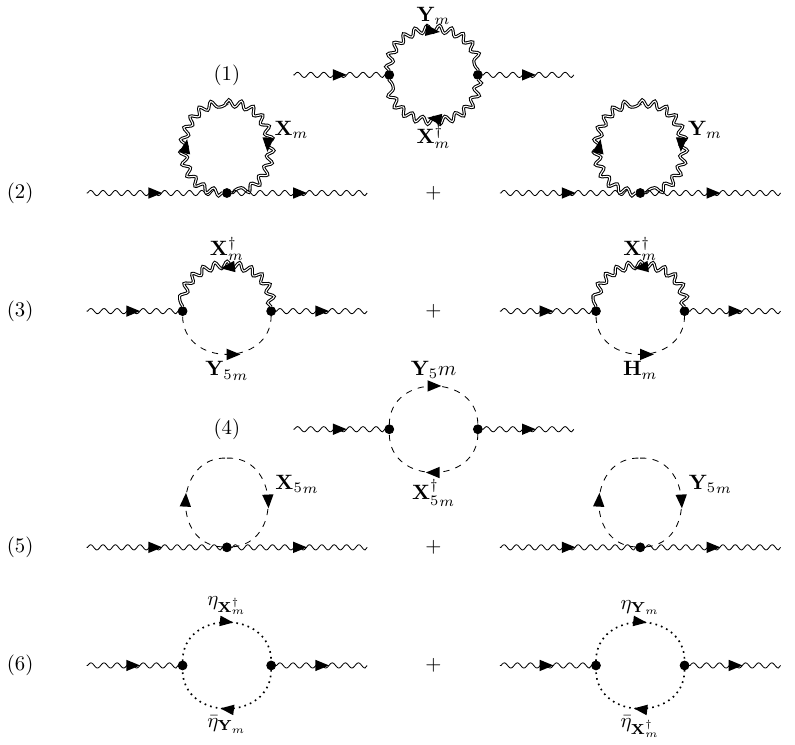}
    \caption{Loops involved in the $W$ self-energy for the new odd subset of loops. The sum on all odd Kaluza-Klein modes denoted $m$ has be to be taken afterward. The associated two point functions can be found Eq.(\ref{AppC:eq8}) }
    \label{fig9}
\end{figure}

\begin{figure}[ht!]
\hspace*{-1.75cm}
    \centering
    \includegraphics[scale=0.9]{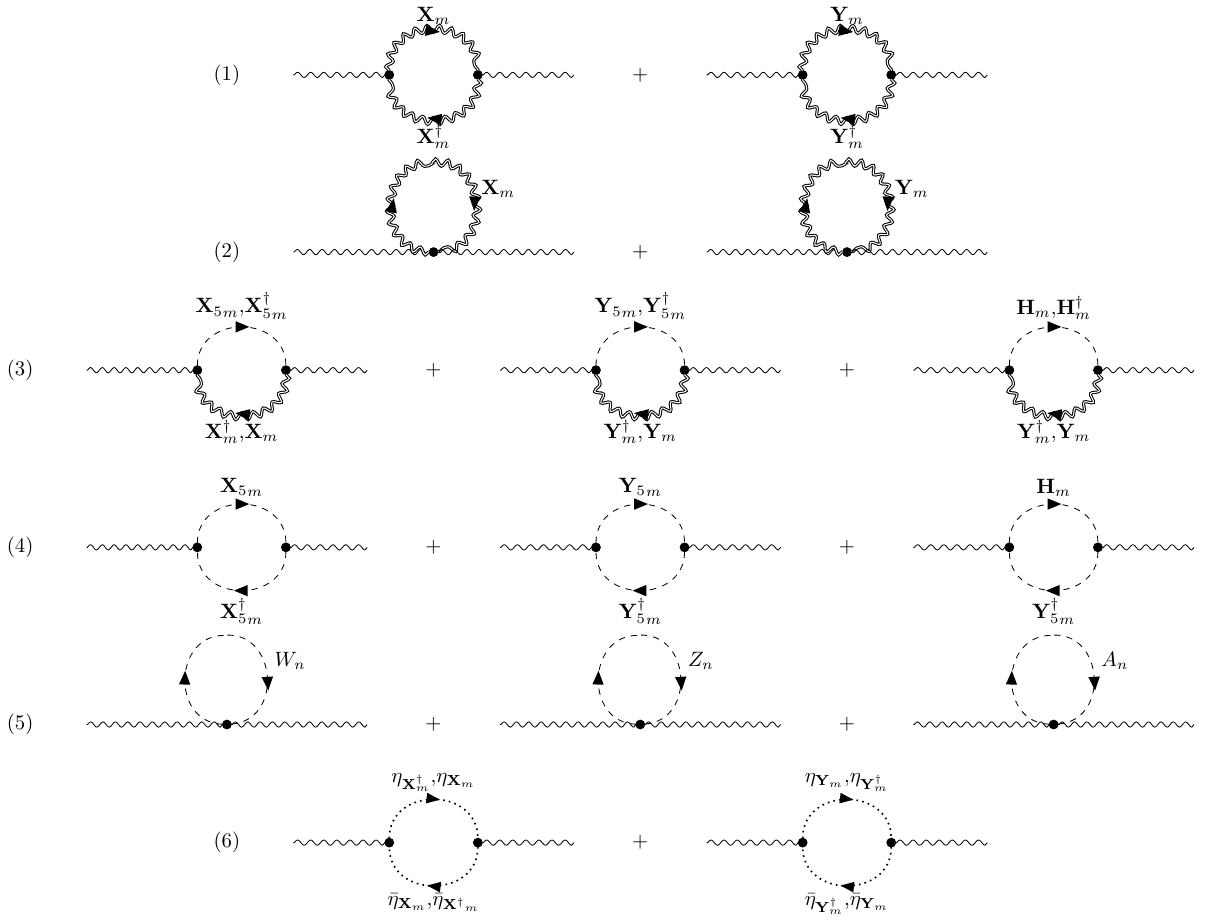}
    \caption{Loops involved in the  $ZZ,\gamma\gamma,Z\gamma$ self-energies for the new-odd subset of loops. The sum on all odd Kaluza-Klein modes denoted $m$ has be to be taken afterward. The associated two point functions can be found Eq.(\ref{AppC:eq9}) }
    \label{fig10}
\end{figure}

\newpage
Finally, we list the loops involving the new odd coloured particles in Figure \ref{fig9} and \ref{fig10}
\newpage
\newpage
Finally, in the same order as before, we associate to each loop of each subset the associated combination of PV function. For the self-energy of the $W$ boson of the SM-like subset, depicted in Figure \ref{fig5}, the functions are 
\begin{equation}
\begin{aligned}
\frac{1}{g^2}\Pi_{WW}^n(p^2 ) = 
1.&\quad c^2_wA^1_1\left(M^Z_n,M^W_n\right)+s^2_wA_1^1\left(M_n,M^W_n\right)\\
2.& - m^2_W B_0(M^h_n,M^W_n)\\
3.&+ A_3^1(M^W_n)+c^2_wA_3^1(M^Z_n)+s^2_wA_3^1(M_n)\\
4.&+\frac{1}{4}A_0(M^h_n)\\
5. &-\frac{s^4_w}{c^2_w}m^2_WB_0(M_n^Z,M_n^W)-s^2_wm_W^2B_0(M_n,M^W_n)\\
6.& +\frac{1}{4}A_0(M_n^Z)+\frac{1}{2}A_0(M_n^W)\\
7.&+ B_{00}(M^W_n,M_n^h)\\
8. &+ B_{00}(M_n^Z,M_n^W)\\
9.&-2c^2_wB_{00}(M_n^Z,M_n^W)-2s^2_wB_{00}(M_n,M_n^W),
\end{aligned}
\label{AppC:eq4}
\end{equation}
and for the $ZZ,\gamma\gamma,Z\gamma$ self-energies Figure \ref{fig6}:
\begin{equation}
\begin{aligned}
\frac{1}{g^2}\Pi_{(ZZ,\gamma\gamma,Z\gamma)}^n(p^2 ) = 
1.&\quad (c^2_w,s^2_w,s_wc_w)A^1_1\left(M_n^W,M_n^W\right)\\
2.& - \left(\frac{m^2_Z}{c^2_w},0,0\right) B_0(M_n^h,M_n^Z)\\
3.&+ 2(c^2_w,s^2_w,s_wc_w)A_3^1(M_n^W)\\
4.&+\left(\frac{1}{4c^2_w},0,0\right)A_0(M_n^h)\\
5. &-2s^2_w\left(\frac{s^2_w}{c^2_w},1,-\frac{s_w}{c_w}\right)m^2_WB_0(M_n^W,M_n^W)\\
6.& \left(\frac{1}{4c^2_w}A_0(M_n^Z)+\frac{(1-2s^2_w)^2}{2c^2_w}A_0(M_n^W),2s^2_wA_0(M_n^W),\frac{s_w}{c_w}(1-2s^2_w)A_0(M_n^W)\right)\\
7.&+ \left(\frac{1}{c_w^2},0,0\right)B_{00}(M_n^Z,M_n^h)\\
8. &+\left(\frac{(1-2s^2_w)^2}{c_w^2},4s^2_w,2\frac{s_w}{c_w}(1-2s^2_w)\right)B_{00}(M_n^W,M_n^W)\\
9.&-2(c^2_w,s^2_w,s_wc_w)B_{00}(M_n^W,M_n^W).
\end{aligned}
\label{AppC:eq5}
\end{equation}
For the new-even subset, only gauge scalar are present thus the number of loop is smaller in comparison with the SM-like subset:
\begin{equation}
\begin{aligned}
\frac{1}{g^2}\Pi_{WW}^{n}(p^2) = 
1.&  \quad s^2_wB_{00}\left(M_n,M_n^W\right)+c^2_wB_{00}\left(M_n^Z,M_n^W\right)\\
2.& \quad +A_0\left(M_n^W\right)\\
3.& \quad +s^2_wA_0\left(M_n\right)+c_w^2A_0\left(M_n^Z\right),
\end{aligned}
\label{AppC:eq6}
\end{equation}
\begin{equation}
\begin{aligned}
\Pi_{(ZZ,\gamma\gamma,Z\gamma)}^{n}(p^2) = 
1.&\quad (c^2_w,s^2_w,s_wc_w)B_{00}\left(M_n^W,M_n^W\right)\\
2.& \quad  2(c_w^2,s_w^2,s_wc_w)A_0\left(M_n^W\right).\\
\end{aligned}
\label{AppC:eq7}
\end{equation}
The odd particles are all coloured, and for simplicity, we always factorise the $3$ factor coming from the colour. The argument of each function is $\left(p^2,M^W_n,M^W_n\right)$, such that we do not specify it later. For the $W$ self-energy Eq.(\ref{AppC:eq8}), results are short, for other EW Eq.(\ref{AppC:eq9}) vectors, since they are in the mass basis and not the $SU(2)\times U(1)_Y$ basis, the results are more cumbersome but are simplified once in the right basis. We recall here that $(I_Y,Q_Y) = (1/2,-1/3)$ and $(I_X,Q_X)=(-1/2,-4/3)$.
\begin{equation}
\begin{aligned}
\frac{1}{3g^2}\Pi_{WW}^{m\epsilon}(p^2) = \quad &1.\quad \frac{1}{2}A_{1}^1\\
&2. +A_{3}^1\\
&3. -\frac{1}{2}\left(M_n^2+m_W^2\right)B_0\\
&4.+2B_{00}\\
&5. + A_0\\
&6.-4B_{00}
\end{aligned}
\label{AppC:eq8}
\end{equation}

\begin{equation}
\begin{aligned}
&\quad\quad\frac{1}{3g^2}\Pi_{ZZ,\gamma\gamma,Z\gamma}^{m\epsilon}(p^2) = \\ &(1)\quad\left[\frac{\left(I_{X,Y}-Q_{X,Y}s^2_w\right)^2}{c^2_w},Q_{X,Y}^2s^2_w,\frac{Q_{X,Y}s_w}{c_w}\left(I_{X,Y}-Q_{X,Y}s^2_w \right) \right]A_1^1 \\
&(2)+2\left[\frac{\left(I_{X,Y}-Q_{X,Y}s^2_w\right)^2}{c^2_w},Q_{X,Y}^2s^2_w,\frac{Q_{X,Y}s_w}{c_w}\left(I_{X,Y}-Q_{X,Y}s^2_w\right)\right]A_3^1\\
&(3)-\left[\frac{\left(I_{X,Y}-Q_{X,Y}s^2_w\right)^2}{c^2_w},Q_{X,Y}^2s^2_w,\frac{Q_{X,Y}s_w}{c_w}\left(I_{X,Y}-Q_{X,Y}s^2_w\right)\right]M_n^2 B_0\\
&-2\left[\frac{\left(I_{Y}-Q_{Y}s^2_w\right)^2}{c^2_w},Q_{Y}^2s^2_w,\frac{Q_{Y}s_w}{c_w}\left(I_{Y}-Q_{Y}s^2_w\right)\right]m_W^2B_0\\
&(4)+4\left[\frac{\left(I_{X,Y}-Q_{X,Y}s^2_w\right)^2}{c^2_w}+Q_H^2\frac{s^4_w}{c^2_w},Q_{X,Y}^2s^2_w+Q_H^2s^2_w,\frac{Q_{X,Y}s_w}{c_w}\left(I_{X,Y}-Q_{X,Y}s^2_w\right)-Q_H^2\frac{s_w^3}{c_w}\right]B_{00}\\
&(5)+2\left[\frac{\left(I_{X,Y}-Q_{X,Y}s^2_w\right)^2}{c^2_w}+Q_H^2\frac{s^4_w}{c^2_w},Q_{X,Y}^2s^2_w+Q_H^2s^2_w,\frac{Q_{X,Y}s_w}{c_w}\left(I_{X,Y}-Q_{X,Y}s^2_w\right)-Q_H^2\frac{s_w^3}{c_w}\right]A_0\\
&(6)-8\left[\frac{\left(I_{X,Y}-2Q_{X,Y}s^2_w\right)^2}{c^2_w},Q_{X,Y}^2s^2_w,\frac{Q_{X,Y}s_w}{c_w}\left(I_{X,Y}-Q_{X,Y}s^2_w\right)\right]B_{00}
\end{aligned}
\label{AppC:eq9}
\end{equation}
\newpage

\section{Propagators and Feynman rules for EW interactions in 't Hoof gauge}\label{App E}
For clarity, we list here the propagator and Feynman rule used for the loops.
Propagators have the usual form in the $R_{\xi}$ gauge, $n$ denote an even number and $m$ an odd number. First, the vector propagators:
\begin{figure}[ht!]
\centering
\includegraphics[scale=1]{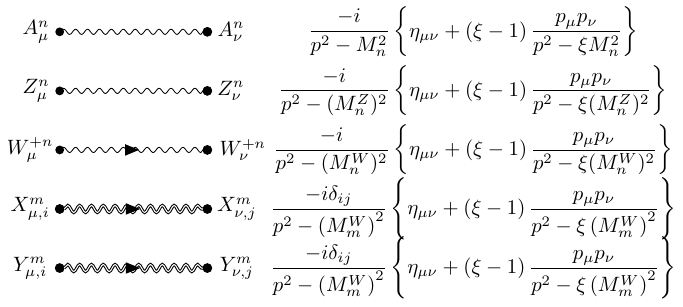},
\end{figure}
\\the scalar propagators
\begin{figure}[ht!]
\centering
\includegraphics[scale=1]{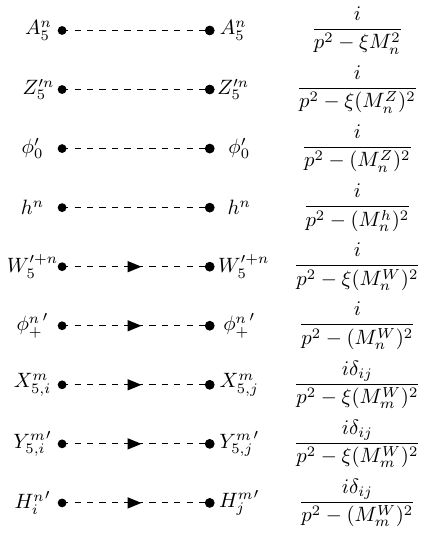},
\end{figure}
\\and the ghost propagators
\begin{figure}[ht!]
\centering
\includegraphics[scale=1]{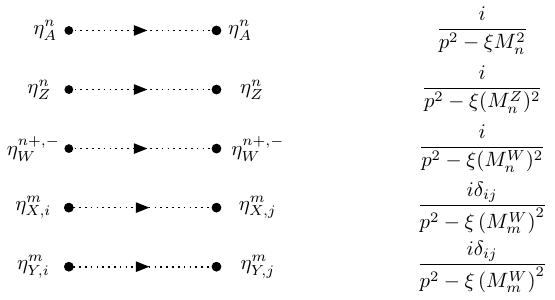}.
\end{figure}

\newpage 
The different vertices are computed with the FeynRules package, the following conventions are used: all momentum are taken inward, no Kaluza-Klein indices means that the Kaluza-Klein number is  and $g_2$ is the gauge coupling of $SU(2)$.
\subsubsection*{Three vectors}
\begin{figure}[ht!]
\centering
\includegraphics[scale=1]{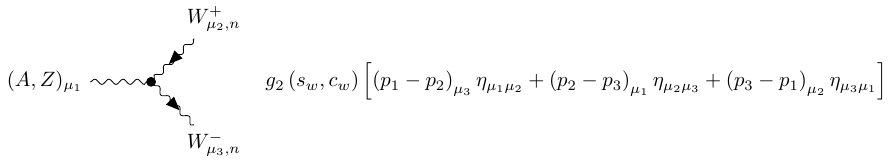}
\end{figure}
\begin{figure}[ht!]
\centering
\includegraphics[scale=1]{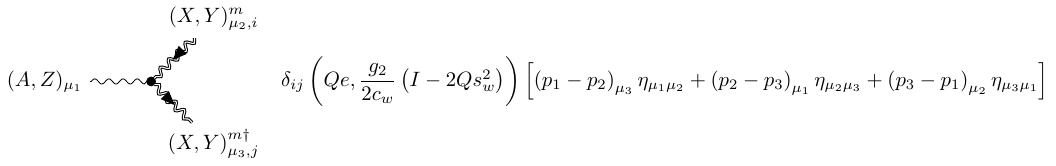}
\end{figure}
\begin{figure}[ht!]
\centering
\includegraphics[scale=1]{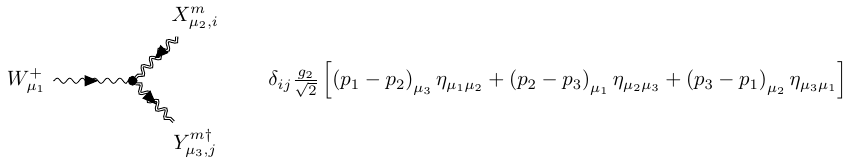}
\end{figure}
\newpage
\subsubsection*{Two vectors one scalar}
\begin{figure}[ht!]
\centering
\includegraphics[scale=1]{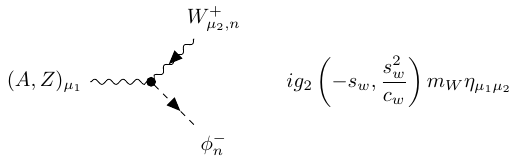}
\includegraphics[scale=1]{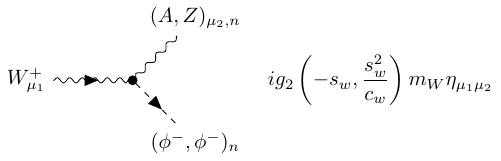}
\end{figure}
\begin{figure}[ht!]
\centering
\includegraphics[scale=1]{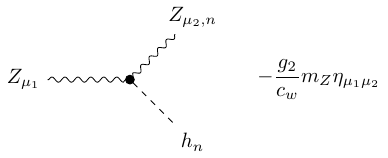}
\includegraphics[scale=1]{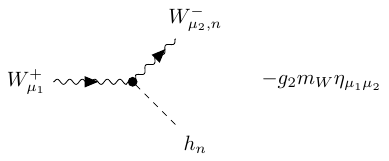}
\end{figure}
\begin{figure}[ht!]
\centering
\includegraphics[scale=1]{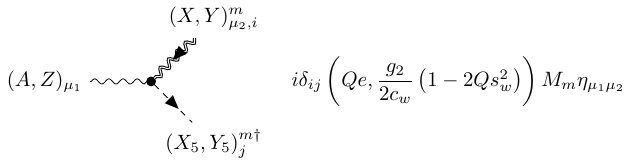}
\end{figure}
\begin{figure}[ht!]
\centering
\includegraphics[scale=1]{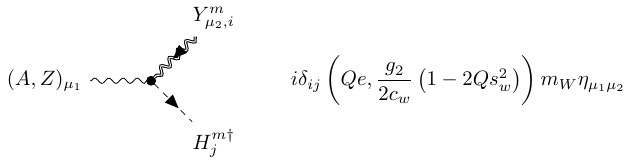}
\end{figure}
\begin{figure}[ht!]
\centering
\includegraphics[scale=1]{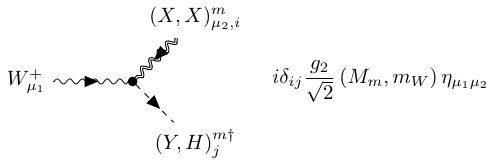}
\end{figure}
\newpage
\subsubsection*{One vector two scalars}
\begin{figure}[ht!]
\centering
\includegraphics[scale=1]{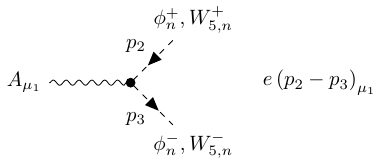}
\includegraphics[scale=1]{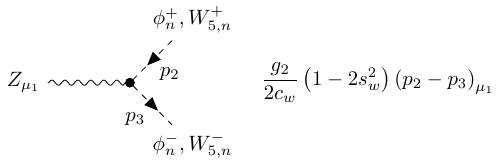}
\end{figure}
\begin{figure}[ht!]
\centering
\includegraphics[scale=1]{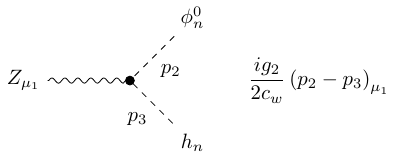}
\includegraphics[scale=1]{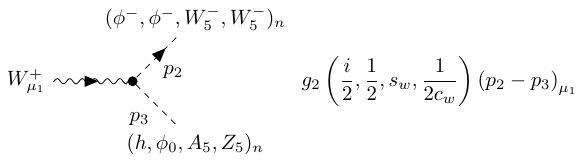}
\end{figure}
\begin{figure}[ht!]
\centering
\includegraphics[scale=1]{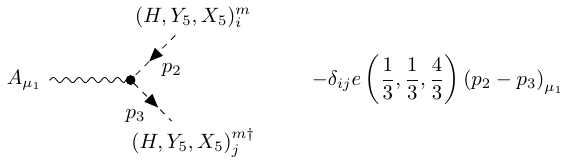}
\includegraphics[scale=1]{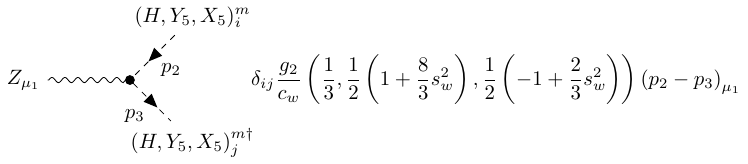}
\end{figure}
\begin{figure}[ht!]
\centering
\includegraphics[scale=1]{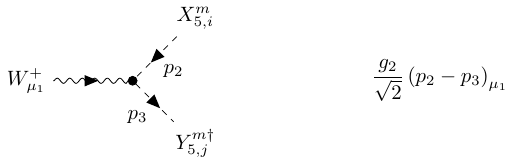}
\end{figure}
\newpage
\subsubsection*{One vector two Fadeev-Popov ghost}
\begin{figure}[ht!]
\centering
\includegraphics[scale=1]{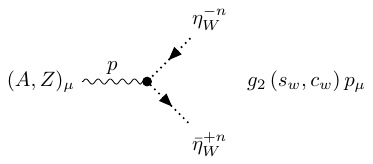}
\includegraphics[scale=1]{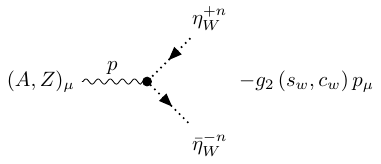}
\end{figure}
\begin{figure}[ht!]
\centering
\includegraphics[scale=1]{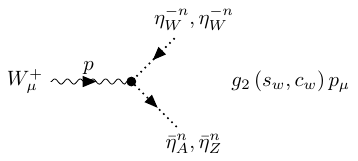}
\includegraphics[scale=1]{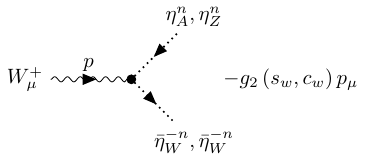}
\end{figure}
\begin{figure}[ht!]
\centering
\includegraphics[scale=1]{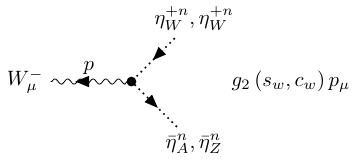}
\includegraphics[scale=1]{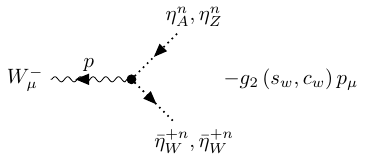}
\end{figure}
\begin{figure}[ht!]
\centering
\includegraphics[scale=0.9]{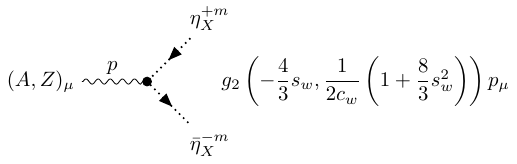}
\includegraphics[scale=0.9]{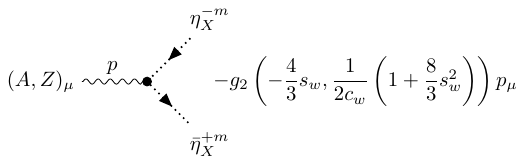}
\end{figure}
\begin{figure}[ht!]
\centering
\includegraphics[scale=0.9]{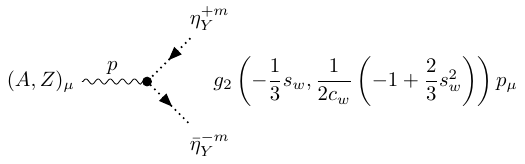}
\includegraphics[scale=0.9]{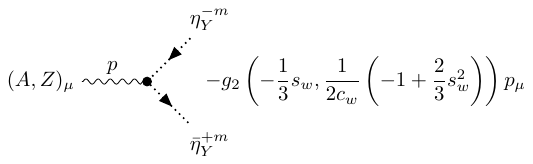}
\end{figure}
\begin{figure}[ht!]
\centering
\includegraphics[scale=1]{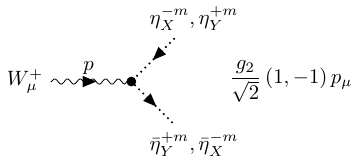}
\includegraphics[scale=1]{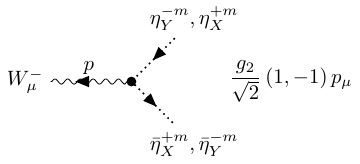}
\end{figure}
\newpage
\subsubsection*{Four vectors}
\begin{figure}[ht!]
\centering
\includegraphics[scale=1]{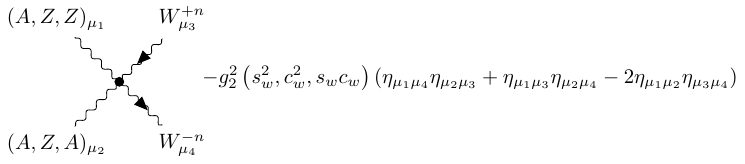}
\end{figure}
\begin{figure}[ht!]
\centering
\includegraphics[scale=1]{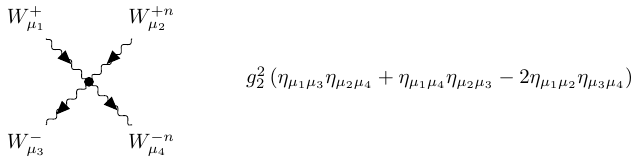}
\end{figure}
\begin{figure}[ht!]
\centering
\includegraphics[scale=1]{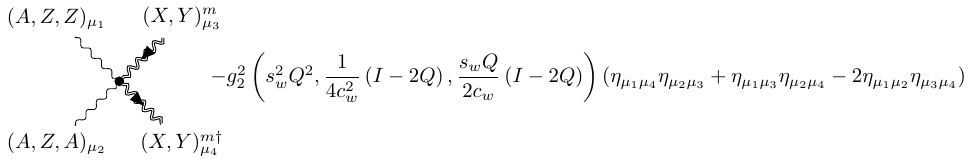}
\end{figure}
\begin{figure}[ht!]
\centering
\includegraphics[scale=1]{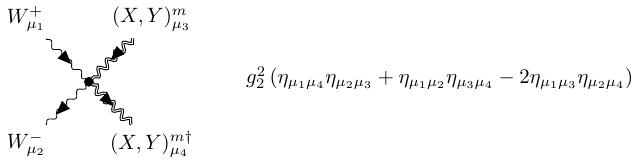}
\end{figure}
\newpage
\subsubsection*{Two vectors two scalars}
\begin{figure}[ht!]
\centering
\includegraphics[scale=1]{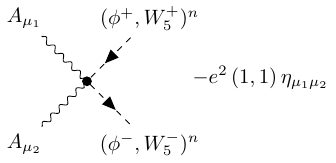}
\includegraphics[scale=1]{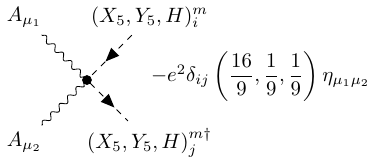}
\end{figure}
\begin{figure}[ht!]
\centering
\includegraphics[scale=1]{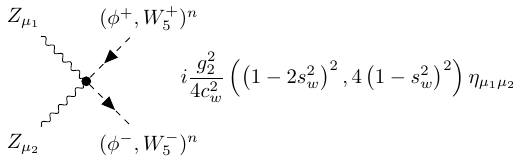}
\includegraphics[scale=1]{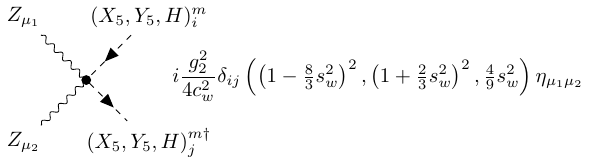}
\end{figure}
\begin{figure}[ht!]
\centering
\includegraphics[scale=1]{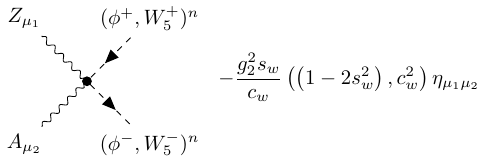}
\includegraphics[scale=1]{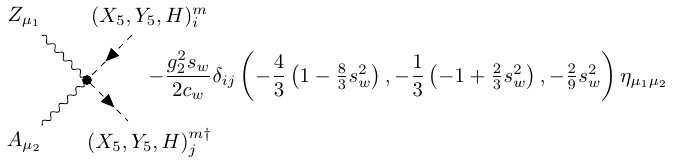}
\end{figure}
\begin{figure}[ht!]
\centering
\includegraphics[scale=1]{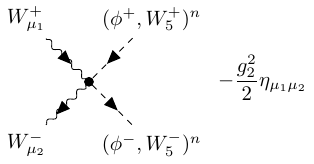}
\includegraphics[scale=1]{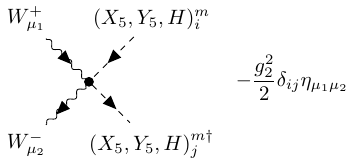}
\end{figure}
\begin{figure}[ht!]
\centering
\includegraphics[scale=1]{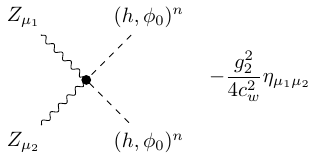}
\includegraphics[scale=1]{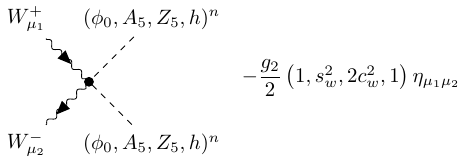}
\end{figure}

\section{$B^2$ production and decay}
 \label{sec:b2}
From the fermionic sector of \ref{sect2.eq1} we extract couplings that involve a 0-mode quark and two odd KK particle, the $10,\overline{10}$ representation give
\begin{equation}
\begin{aligned}
& \Tr\,\left(\overline{\Psi}_{10_L}i\slashed{D}\Psi_{10_L}\right)\supset g_S\overline{T}_L^c\slashed{X}q_{t_L} + g_S\overline{T}_L^c\slashed{Y}q_{b_L} + h.c \\
& \Tr\,\left(\overline{\Psi}_{\overline{10}_L}i\slashed{D}\Psi_{\overline{10}_L}\right)\supset g_S\overline{t}_R\slashed{X}^{\dagger}Q_{T_R}^c + g_S\overline{t}_R\slashed{Y}Q_{B_R}^c + h.c \\
& \quad \quad\quad\quad\quad\quad\quad\quad\: +\: g_5\overline{Q}_{T_R}\slashed{Y}\tau_R^c + g_5\overline{Q}_{B_R}\slashed{X}\tau_R^c + h.c,
\end{aligned}
\end{equation}
and the $5,\overline{5}$ 
\begin{equation}
\begin{aligned}
&\overline{\Psi}_{\overline{5}}i\slashed{D}\Psi_{\overline{5}} \supset g_5\overline{B}^c_L\slashed{X}l_{\tau_L}+g_5\overline{B}^c_L\slashed{Y}l_{\nu_L}h.c\\
& \overline{\Psi}_{5}i\slashed{D}\Psi_{5} \supset g_5\overline{b}_R\slashed{X}L_{{\mathcal{T}}_R}+g_5\overline{b}_R\slashed{Y}L_{N_R}h.c,\\
\end{aligned}
\end{equation}
where 
\begin{equation}
    \slashed{A} = \eta_{MN}\gamma_NA_M \qq{with} \gamma_M = \{\gamma_{\mu}, -i\gamma_5\}.
\end{equation}
Couplings with the scalar sector are also to be taken into account
\begin{equation}
\begin{aligned}
&-Y_b\sqrt{2}\Psi_5\Psi_{10}\phi_5^*\supset -Y_b\left(\bar{b}_RH^*T_R^c+\overline{L}_{\mathcal{T}_L}^cH^*q_{t_L}+\overline{L}_{N_L}^cH^*q_{b_L} \right) +h.c\\
&-Y_{\tau}\sqrt{2}\overline{\Psi}_{\overline{10}}\Psi_{\overline{5}}\phi_5^*\supset-Y_{\tau}\left(\bar{t}_RH^*B^c_R +\overline{Q}_{T_L}H^*l_{\tau_L}+\overline{Q}_{B_L}H^*l_{\nu_L}\right) +h.c\\
&-\frac{1}{2}Y_t\epsilon_5\overline{\Psi}_{\overline{10}}\Psi_{10}\phi_5 \supset -Y_t\left(\overline{Q}^c_{B_L}Hq_{t_L}+\overline{Q}^c_{T_L}Hq_{b_L}+\bar{t}_RH\mathcal{T}_R^c+\overline{T}^c_RH^*\tau_R\right)+h.c
\end{aligned}
\end{equation}

\begin{figure}[h!]
    \centering
    \includegraphics[width=0.23\linewidth]{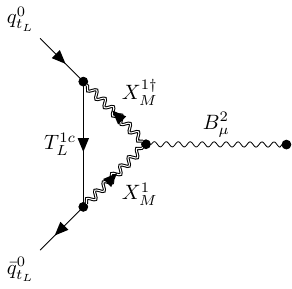}
    \includegraphics[width=0.23\linewidth]{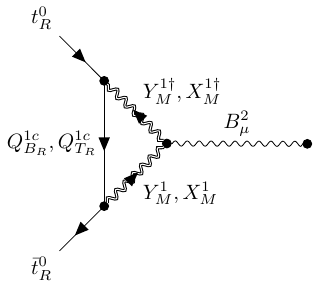}
    \includegraphics[width=0.23\linewidth]{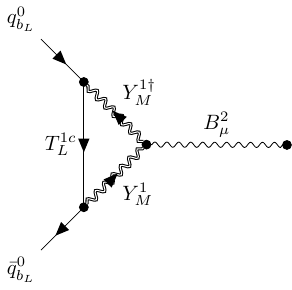}
    \includegraphics[width=0.23\linewidth]{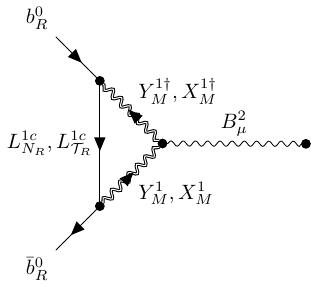}
    \includegraphics[width=0.23\linewidth]{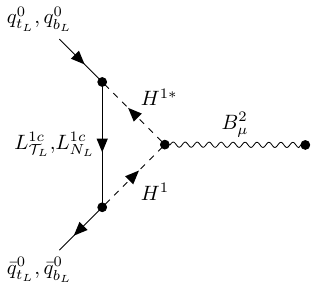}
    \includegraphics[width=0.23\linewidth]{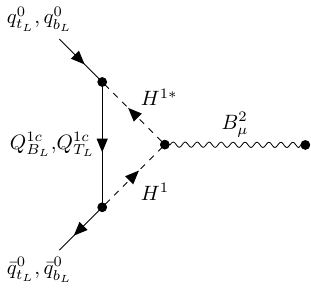}
    \includegraphics[width=0.23\linewidth]{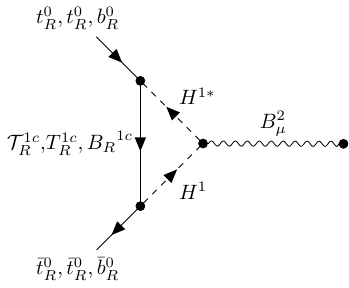}
    \caption{Channel for the production of the 2-mode $B^2_{\mu}$ from $q\bar{q}$ initial states through loop of 1-mode. In the case of the production of $g^2$ the loops of 1-mode are the same.}
    \label{sect4:fig4}
\end{figure}

\begin{figure}[h!]
    \centering
    \includegraphics[width=0.23\linewidth]{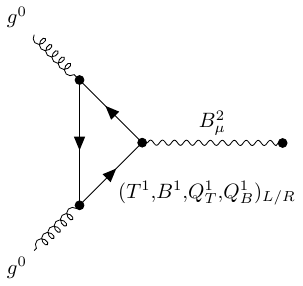}
    \includegraphics[width=0.23\linewidth]{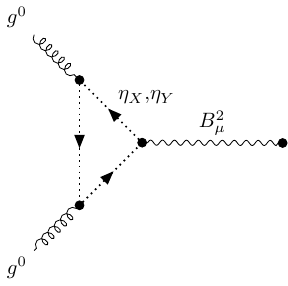}
    \includegraphics[width=0.23\linewidth]{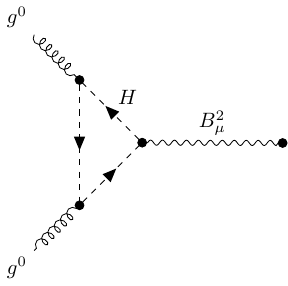}
    \includegraphics[width=0.23\linewidth]{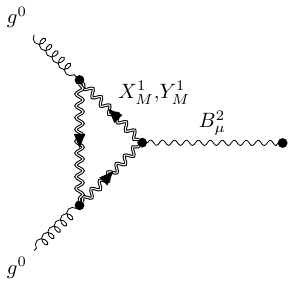}
    \caption{Channel for the production of the 2-mode $B^2_{\mu}$ from $gg$ initial states through loop of 1-mode. In the case of the production of $g^2$ the loops of 1-mode are the same.}
    \label{sect4:fig5}
\end{figure}

\begin{figure}[h!]
    \centering
    \includegraphics[width=0.23\linewidth]{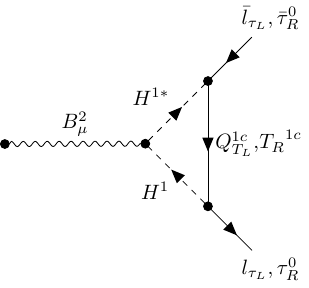}
    \includegraphics[width=0.23\linewidth]{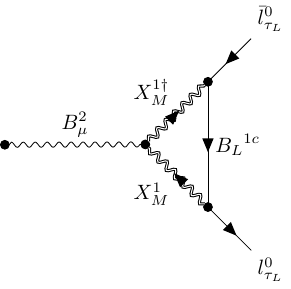}
    \includegraphics[width=0.23\linewidth]{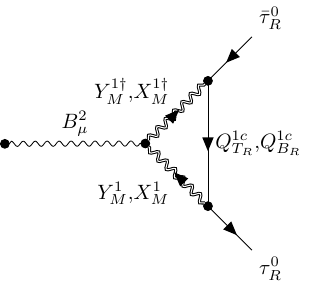}
    \caption{Decay of the 2-mode $B^2_{\mu}$ through different loops of 1-mode.}
    \label{sect4:fig6}
\end{figure}

\newpage
\bibliographystyle{utphys}
\bibliography{aGUT}
\end{document}